\documentclass[11pt]{article}
\usepackage{graphicx,amsmath,amssymb,epsf, hyperref} 
\usepackage{latexsym,bm,slashed} 
\usepackage{xcolor} \definecolor{dark}{rgb}{0.10,0.2,0.3}
\definecolor{magenta}{rgb}{0.7,0.1,0.3}
\definecolor{purpure}{rgb}{0.5,0.15,0.3}
\usepackage[font=small,format=plain,labelfont=bf,up,textfont=it,up]{caption}
\usepackage{hyperref, cite} \hypersetup{colorlinks, citecolor=blue,
  filecolor=blue, linkcolor=magenta,
  urlcolor=purpure,hyperfootnotes=true,pdftex} 

\newcommand{\kt}{{\bm k}}

\newcommand{\rt}{{\bm r}}

\newcommand{\asb}{\bar{\alpha}_s}
 \title{\bf \Large The energy dependence of exclusive heavy vector meson photoproduction cross-sections  and  NLO  BFKL evolution }
 \author{Martin Hentschinski and Ricardo Rangel Ram\'irez\\ \\ 
Departamento de Actuaria, F\'isica y Matem\'aticas,
Universidad de las Americas Puebla, \\ Santa Catarina Martir, 72820 Puebla, Mexico }

\begin{document}

\maketitle
\begin{abstract}
We study the energy dependence of the cross-section for exclusive photoproduction of vector mesons $J/\psi$ and $\Upsilon$, using a solution to the next-to-leading order (NLO)  Balitsky-Fadin-Kuraev-Lipatov (BFKL) equation. Our goal is to use BFKL evolution as a benchmark  to provide evidence for the presence of non-linear QCD dynamics and signs for the onset of gluon saturation at highest center of mass energies. Our approach determines initial conditions  for the  proton from the  Bartels-Golec Biernat-Kowalski (BGK) dipole model, and  evolves the resulting unintegrated gluon distribution using NLO BFKL evolution. For the nucleus, initial conditions are generated through both the impact parameter dependent saturation model (IP-Sat), using a Wood-Saxon distribution, and  a $A^{1/3}$ scaling of the saturation scale of the original BGK model.  We find that NLO BFKL evolution provides a very good description of the nuclear modification factor for $J/\psi$ production, if initial conditions are generated through a  $A^{1/3}$ scaled BGK model, while the description fails, if initial conditions are created using the IP-Sat model.
 \end{abstract}

\section{Introduction}
\label{sec:intro}

Exclusive photoproduction of vector mesons in ultraperipheral
collisions at the Large Hadron Collider (LHC) is a very useful process  to explore Quantum Chromodynamics (QCD) at
ultra-small values of $x$. Here, $x = M_V^2/W^2$, with $M_V$  the
mass of the vector meson and $W$  the center of mass energy of the
photon-proton or photon-nucleus reaction. The process is primarily
of interest for the photoproduction of charmonium, where the charm mass
provides a perturbative scale at the boundary to non-perturbative
dynamics; it is therefore suitable to potentially observe effects related to the
saturation of gluon densities at ultra-low values of $x$
\cite{Gribov:1984tu,Morreale:2021pnn,Hentschinski:2022xnd,Aguilar:2024otb}. 
At low $x$, perturbative QCD predicts through the
Balitsky-Fadin-Kuraev-Lipatov (BFKL)  equation
\cite{Kuraev:1977fs,Kuraev:1976ge, Balitsky:1978ic} a powerlike rise
of the gluon distribution. Bounds due to unitarity state however that
such a rise cannot continue to arbitrarily small values of $x$. At some value of $x$ it must start to  slow down and eventually come to hold.  A
theoretical framework which predicts such a slow down and saturation of gluon densities are
the JIMWLK-BK equations \cite{Balitsky:1995ub,
  Jalilian-Marian:1997ubg, Kovchegov:1999yj, Iancu:2000hn,
  Weigert:2000gi, Iancu:2001ad, Ferreiro:2001qy}, which generalize
BFKL evolution to the case of high gluon densities. 

Photoproduction of charmed vector mesons in
exclusive reactions has been extensively explored 
during the recent years, see e.g. \cite{Bautista:2016xnp, Cepila:2018faq,Garcia:2019tne,
  Krelina:2019gee, Klein:2019qfb, Kopeliovich:2020has,
  Bendova:2020hbb,Hentschinski:2020yfm, Jenkovszky:2021sis, Flett:2021xsl,
  Mantysaari:2021ryb, Mantysaari:2022kdm, Goncalves:2022ret,
  Wang:2022vhr, Mantysaari:2023xcu,Cepila:2023dxn,Mantysaari:2024zxq, Cepila:2024qge, Penttala:2024hvp, Cepila:2025rkn, Nemchik:2025myg,Goncalves:2024jlx} for recent theory and \cite{Klein:2020nvu,
  Bylinkin:2022wkm, ALICE:2023fov,daCosta:2025frd} for experimental proposals as well
as the reviews \cite{Amoroso:2022eow,
  Frankfurt:2022jns,Arleo:2025oos}. If data sets collected at both HERA and
LHC experiments are combined, such reactions allow to explore low $x$ dynamics over several orders of magnitude, from
$x\simeq 0.1$ down to $x =10^{-6}$. Exclusive photoproduction on  large nuclei is of
special interest since  gluon densities
are in this case further enhanced through the nuclear mass number $A$. Nevertheless,
comparison of recently measured ALICE and CMS data for photonuclear
production on lead \cite{ALICE:2023jgu,CMS:2023snh} with theory
predictions do not yet allow to draw firm conclusions on the  observation of gluon saturation effects. In particular, descriptions
based on a leading twist approach, using nuclear modification factors
of parton distribution functions (PDFs), provide currently a competitive description, albeit $x$-dependence characteristic for gluon saturation close to the DGLAP starting scale might be hidden in this case in the initial conditions.

A question which one therefore would like to answer in such reactions is whether
the observed energy dependence can be explained through an approach
which assumes low gluon densities or whether inclusion of high density
effects is needed. Predictions which include high density effects are usually
based on solutions of non-linear QCD evolution equations, such as the
JIMWLK-BK evolution equations. Nevertheless the ability of such
schemes to describe data is insufficient to provide evidence for high
gluon densities and potentially gluon saturation, since the evolution equations itself are also applicable in the  low density regime. 
One therefore 
needs in addition a framework which assumes low gluon densities and
which fails to describe the data set. At first sight, a natural framework for the low density limit is  collinear
factorization, which provides predictions not only for the low $x$
region but for all values of $x$, while non-perturbative input to
Dokshitzer-Gribov-Lipatov-Altarelli-Parisi (DGLAP) evolution equations
can be constrained by global fits. While appealing for the
aforementioned reasons, relying on collinear factorization only bears the
danger to overlook the presence of high gluon densities at low hard scales.   DGLAP evolution  evolves PDFs from low to high hard scales and testing its validity requires  two reactions with a hierarchy of hard scales, i.e.  photoproduction of the $J/\psi$ (low hard scale) and the $\Upsilon$ (high hard scale). Since $x= M_V^2/W^2$ and with the maximal center of mass energy $W$ fixed, bottonium is limited to larger $x$ values than charmonium production,  $x_{\Upsilon}>x_{J/\psi}$. In addition,  predictions at the bottom scale probe  only  initial conditions at the charm scale for $x\geq x_{\Upsilon}$. Parametrizations at $x< x_\Upsilon$ do not enter the  DGLAP evolution.   Predictions at low hard scales and low $x<x_{\Upsilon}$,
such as $J/\psi$ photoproduction , therefore either depend strongly on
extrapolation of the parametrization used at larger values of $x$ or
-- if $J/\psi$ data are included in the fit -- fit merely the
observed $x$ dependence.

To provide at the very least an alternative framework to test the validity of a low density approach,  one may therefore use BFKL evolution to evolve hadronic structure  at intermediate $x \simeq 0.01$ down to low $x$.  If the BFKL prediction significantly deviates from observation, one can  quantify the degree of deviation and therefore arrive at a conclusion to which degree the low density description breaks down. Note that such an approach has been taken before in the literature, see in particular \cite{Garcia:2019tne,Hentschinski:2020yfm,Penttala:2024hvp}.
Here  \cite{Garcia:2019tne,Hentschinski:2020yfm} rely on a fit of the proton impact factor to HERA data and therefore are naturally limited to photoproduction on a proton. \cite{Penttala:2024hvp}  provides on other hand  predictions for $J/\psi$ and $\Upsilon$ photoproduction on both proton and a lead nucleus. While the scope of the present study and \cite{Penttala:2024hvp} is very similar, there exist a few important differences which we would like to highlight in the following.  \cite{Penttala:2024hvp} provides predictions for both BK (non-linear) and BFKL (linear) evolution equations based on the leading order BFKL kernel, supplemented with NLO running coupling corrections; in addition  \cite{Penttala:2024hvp} uses a gluon mass to regulate a certain infrared instability associated with impact parameter dependence in the evolution equation and is therefore capable to provide predictions for finite  momentum transfer $t$ between photon and proton/lead nucleus. Our treatment is on the other hand restricted to  BFKL evolution only, while we will take  into account the complete NLO corrections, including a resummation of configurations enhanced by collinear logarithms, which have been found to provide important contributions for a correct description of data, see e.g. \cite{Hentschinski:2012kr,Hentschinski:2013id}. The  solution to the NLO BFKL evolution equation is however restricted to zero momentum transfer.

Apart from  potential uncertainties related to the restriction to $t=0$,  this approach is subject to  uncertainties due to the  transverse momentum distribution of the proton or lead  impact factor  at the initial $x$ value as well as on the particular choice which one takes to solve the  NLO  BFKL evolution equations.  To constrain the initial transverse momentum distribution for the proton, we will use, unlike  \cite{Bautista:2016xnp,Garcia:2019tne,Hentschinski:2020yfm}, the recently refitted  \cite{Golec-Biernat:2017lfv} Bartels-Golec Biernat-Kowalski (BGK) model \cite{Bartels:2002cj} to generate an unintegrated gluon densities at $x= 0.01$. For the lead nucleus, we use the impact parameter dependent saturation model \cite{Kowalski:2003hm}, based on the BGK model, as well as the BGK model with the  effective saturation scale  re-scaled by a factor  $A^{1/3}$. Since study of these models is of interest in its own right, we will provide in the following also a brief comparison of the energy dependence of these models to data, while our  main interest   is in  using them as a  way to model initial conditions for NLO BFKL evolution. 
Our solution of the NLO BFKL equation follows closely the solution of  \cite{Hentschinski:2012kr,Hentschinski:2013id}, which has been tested both in a  fit to HERA data as well as LHC phenomenology \cite{Chachamis:2015ona, Celiberto:2018muu, Bautista:2016xnp, Garcia:2019tne, Hentschinski:2020yfm}.  The important difference to these studies is that instead of fitting the initial transverse momentum distribution to HERA data, we will use an already existing dipole model, from which we extract the required impact factor. 

To test this setup, we will first explore the proton case, including a comparison to  both $J/\psi$ and $\Upsilon(1s)$  data. Comparison with the $\Upsilon$ photoproduction cross-section provides a test in a region of phase space where BFKL evolution  is expected to perform well.  $J/\psi$ photoproduction takes on the other hand  place at the boundary between perturbative and non-perturbative physics and associated uncertainties are naturally large. Of particular interest is here the study of the nuclear modification factor, since for this quantity uncertainties present in both proton and lead cross-sections will cancel to a large extend.

The outline of this paper is as follows: In Sec.~\ref{sec:setup} we summarize the setup which allows us to obtain the energy dependence of photoproduction cross-sections from the inclusive unintegrated gluon distribution. We also compare the energy dependence predicted by the BGK model and its nuclear generalizations to data. Sec.~\ref{sec:bfkl} discusses the implementation of the solution of the  NLO BFKL equation. Sec.~\ref{sec:num} contains our results, including a comparison to data.  Sec.~\ref{sec:concl} provides our conclusions.

\section{Photoproduction cross-sections}
\label{sec:setup}

\begin{figure}[t]
  \centering
   \includegraphics[width = .5\textwidth]{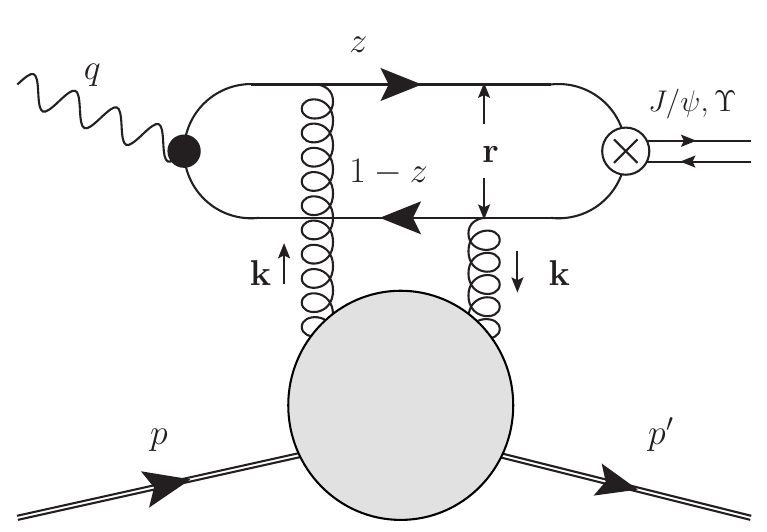}
  \caption{Exclusive photoproduction of vector mesons $J/\psi$ and
    $\Upsilon(1s)$. For the quark-anti quark dipole, we indicate photon momentum
    fractions $z$ and $1-z$ as well as the transverse separation ${\bm
    r}$. Finally ${\bm k}$ denotes the transverse momentum transmitted
from the unintegrated gluon distribution of the proton or lead nucleus; the latter is indicated
through the gray blob; ${\bm r}$ denotes the transverse size of the dipole.}
  \label{fig:reaction}
\end{figure}

We  study the process, 
\begin{align}
  \label{eq:30}
 \gamma(q) + N(p) & \to V(q') + N(p'), & V &= J/\psi, \Upsilon(1s) , \,  N = p, Pb\, ,
\end{align}
where  $\gamma$ denotes a quasi-real
photon with virtuality $Q \simeq 0$. $W^2 = (q + p)^2$ is the 
center-of-mass energy squared of the $\gamma(q) + N(p)$ reaction, see also Fig.~\ref{fig:reaction}.  With  $t = (q - q')^2$, the differential cross-section for
the exclusive photo-production of a vector meson can be written as
\begin{align}
  \label{eq:16}
  \frac{d \sigma}{d t} \left(\gamma N \to V N \right) 
& = 
\frac{1}{16 \pi} \left|\mathcal{A}_{T}^{\gamma p \to V p}(x, t) \right|^2  , & V &= J/\psi, \Upsilon(1s) \, ,
\end{align}
with $\mathcal{A}_T(x, t)$  the scattering amplitude for a transverse polarized real photon with color singlet exchange in the
$t$-channel, where an overall factor $W^2$ has been already extracted.  $x = M_V^2/W^2$ with $M_V$ the mass of the vector meson; see also  { e.g.} \cite{Bautista:2016xnp}.  To access the inclusive gluon distribution, we 
follow a two-step procedure, frequently employed in the
literature:  First  one determines the differential cross-section at zero
momentum transfer $t=0$, which can be related to  the
inclusive gluon distribution; in  a second step  one uses the ansatz  $d\sigma/dt \sim  \exp\left[-|t| B_D\right]$ with the diffractive slope parameter $B_D$, which is determined from data. The total cross-section  then reads
\begin{align}
  \label{eq:16total}
 \sigma^{\gamma p \to V p}(W^2) & = \frac{1}{B_D(W)} \frac{d \sigma}{d t} \left(\gamma p \to V p \right)\bigg|_{t=0}
.
\end{align}
For photon-proton collisions, we use
\begin{align}
  \label{eq:18}
  B_{D,p}^{J/\psi}(W) & =\left[  b_{0,p} + 4 \alpha'_{p} \ln \frac{W}{W_0} \right] \text{GeV}^{-2},
\end{align}
where  $W_0 = 90$~GeV and  $b_{0,p}$ and $\alpha'_p$ have been determined in \cite{Cepila:2019skb} from a fit to HERA data with $b_{0,p}^{J/\Psi} = 4.62$, while $\alpha_p^{'J/\Psi} = 0.171$. For the $\Upsilon$ we use $ B_{D,p}^{ \Upsilon}(W) =  B_{D,p}^{J/\psi}(W) - b_1 \ln (m_\Upsilon^2/m_{J/\psi}^2)$ \cite{Cepila:2019skb}  with $b_1 = 0.45$~GeV$^{-2}$ and $m_\Upsilon = 9.46040$~GeV while $m_{J/\psi} =  3.096916$~GeV. 
For photonuclear $J/\Psi$ production the value  $B_{D, \text{Pb}} =  (4.01\pm 0.15 )\cdot 10^{2}/$GeV$^2$ has been determined in \cite{Peredo:2023oym} from a fit to  ALICE data  \cite{ALICE:2021tyx}, which we will also use for $\Upsilon$ photoproduction. The specific form of Eq.~\eqref{eq:18} indicates a relatively weak energy dependence for the diffractive slope and we therefore do not expect that the above approximations significantly affect our theoretical predictions. Nevertheless our predictions rely on an extrapolation of this parameter at low center of mass energies and a drastic change of its energy dependence at higher energies cannot be excluded with certainty.

The subleading real part of the scattering amplitude
can be  estimated  from the imaginary part using
\begin{align}
  \label{eq:32}
  \frac{\Re\text{e} \mathcal{A}(x)}{\Im\text{m} \mathcal{A}(x)}
&=
\tan \frac{\lambda(x) \pi }{2},
&
 \lambda(x) & = 
\frac{d \ln \Im\text{m}  \mathcal{A}(x) }{ d \ln 1/x} \, .
\end{align}
As noted in \cite{Bautista:2016xnp}, the dependence of the slope
parameter $\lambda$ on $x$ provides a sizable correction to
the  $W$ dependence of the complete
cross-section. We therefore do not assume $\lambda =$const., but determine the slope $\lambda$ directly from the $x$-dependent imaginary part of the scattering amplitude.

\subsection{Wave function overlap}
\label{sec:WFO}

Within high energy factorization, the imaginary part of the scattering amplitude at $t=0$ is obtained as a convolution of the light-front wave function overlap  and the dipole cross-section.   Using  a simple boosted Gaussian model for the vector meson  wave function,  based on the  Brodsky-Huang-Lepage prescription
\cite{Brodsky:1980vj, Cox:2009ag, Nemchik:1994fp}, we have
\begin{align}
\label{am-i}  
\Im \text{m}\mathcal{A}^{\gamma p\rightarrow
    Vp}_{T} (x, t=0) &= \int\!d^2{\bm r} \int_0^1 \frac{dz}{4 \pi}\,
 \left(\Psi_{V}^{*}\Psi_T\right)(r,z) 
   \,
  \sigma_{q\bar{q}}\left(x,r\right)\,, 
\end{align}
where $r = |{\bm r}|$ denotes the transverse separation of the dipole and \cite{Kowalski:2006hc}
\begin{align}
  \label{eq:21}
 \left(\Psi_{V}^{*}\Psi_T\right)  &= \frac{\hat{e}_f eN_c}{\pi z (1-z)}
 \bigg\{
 m_f^2 K_0(m_f r) \phi_T(r,z) - \left[z^2 + (1-z)^2 \right] m_f K_1(m_f r) \partial_r \phi_T(r,z)
\bigg\}
 \, ,
\end{align}
with
\begin{align}
\label{eq:1s2s_groundstate}
\phi_{T}^{1s}(r,z) &= \mathcal{N}_{T, 1s} z(1-z)
  \exp\left(-\frac{m_f^2 \mathcal{R}_{1s}^2}{8z(1-z)} - \frac{2z(1-z)r^2}{\mathcal{R}_{1s}^2} + \frac{m_f^2\mathcal{R}_{1s}^2}{2}\right)  \, .
\end{align}
The free parameters of Eq.~\eqref{eq:1s2s_groundstate} have been
determined in various studies from the  normalization and orthogonality of the wave functions as well as  the decay width of the vector mesons. In the
following we use the values found in 
\cite{Armesto:2014sma,Goncalves:2014swa}  which we summarize in Tab.~\ref{vm_fit}.
\begin{table}
\centering
\caption{Parameters of the boosted Gaussian vector meson wave functions for $J/\Psi$   \cite{Armesto:2014sma} and $\Upsilon(1s)$  \cite{Goncalves:2014swa}.}

\begin{tabular}{c|c|c|c|c|c}
\hline\hline &&&& \vspace{-.2cm}\\
Meson & $m_f/\text{GeV}$  & $\mathcal{N}_T$ &  $\mathcal{R}^2$/$\text{GeV}^{-2}$
& $M_V$/GeV   
\\
&&&&& \vspace{-.2cm}\\  \hline
$J/\Psi$ & $m_c=1.4$&   $0.57$ & $2.45$   &$3.097$      \\ \hline
$\Upsilon(1s)$ & $m_b = 4.2$ & $0.481$ & $0.57$ & $9.460$   \\ \hline
\end{tabular}
\label{vm_fit}
\end{table}

\subsection{Dipole cross-sections}
\label{sec:gluon}

A  simple model for the dipole cross-section is provided by the Bartels, Golec-Biernat, Kowalski (BGK) model \cite{Bartels:2002cj},
\begin{align}
  \label{eq:sigBGK}
  \sigma_{q\bar{q}}^{\text{BGK}}(x, r) & = \sigma_0^{\text{BGK}}  \left[ 1 -  \exp \left(-\frac{r^2 \pi^2\alpha_s(\mu_r^2) xg(x, \mu_r^2) }{3 \sigma_0^{\text{BGK}} }  \right)\right],
\end{align}
which is obtained through exponentiating the collinear dipole cross-section  \cite{Frankfurt:1996ri, Bartels:2002cj}. The model has  been recently refitted \cite{Golec-Biernat:2017lfv} for dipole scattering on a proton  to combined HERA data.  The collinear gluon distribution $g(x, \mu^2)$ is  subject to a leading order DGLAP evolution equation without quarks,
\begin{align}
  \label{eq:dglap}
  \frac{d}{d\mu^2} g(x, \mu^2) & = \frac{\alpha_s}{2 \pi} \int_x^1\frac{dz}{z} P_{gg}(z) g(x/z, \mu^2), & xg(x, Q_0^2) & = A_g x^{-\lambda_g} (1-x)^{5.6},
\end{align}
where $xg(x, Q_0^2)$ denotes the gluon distribution at the initial scale $Q_0 = 1$~GeV. Following  \cite{Golec-Biernat:2017lfv}, we evaluate the  gluon distribution  and the QCD running coupling at the dipole size dependent scale
\begin{align}
  \label{eq:muR}
  \mu_r^2 & = \frac{\mu_0^2}{1 - \exp\left(- \mu_0^2 r^2/C \right)}.
\end{align}
The remaining parameters of the model have been obtained as    $\sigma_0^{\text{BGK}} = (22.93 \pm 0.27)$~mb,  $A_g = 1.07 \pm 0.13$,  $\lambda_g = 0.11 \pm 0.03$,  $C = 0.27 \pm 0.04$,  $\mu_0^2 = (1.74 \pm 0.16 )$~GeV$^2$.   To arrive at a suitable generalization of the BGK model for dipole scattering on a large nucleus, we consider  two  options: the impact parameter dependent saturation (IPSat) model  and a description where the effective saturation scale $Q_s^2(x)$ is scaled by a factor $A^{\frac{1}{3}}$ with $A$ the number of nucleons in the nucleus. The formulation of the IPSat model is based on the optical Glauber model.  Using a Wood Saxon distribution,  one averages over the position of different nucleons  and finds  \cite{Kowalski:2003hm,Mantysaari:2018nng}
\begin{align}
  \label{eq:dsigWS}
  \frac{d\sigma_{q\bar{q}}^{\text{IPSat}}}{d^2 b} & = 2 \left[1-\left(1- \frac{1}{2}T_A(b)\sigma_{q\bar{q}}(x, r) \right)^A \right],
\end{align}
with nuclear thickness function $T(b)$ and Wood-Saxon distribution $\rho^{\text{WS}}(r)$ given by  
\begin{align}
  \label{eq:Tb}
  T(b) & = \int_{-\infty}^\infty dz \rho^{\text{WS}}(\sqrt{z^2 + b^2}),
&
 \rho^{\text{WS}}(r) & = \frac{\mathcal{N}^{\text{WS}}}{1 + e^{\frac{r - R_{Pb}}{d}}}; 
\end{align}
 $\mathcal{N}^{\text{WS}}$ is fixed through  $\int d^3 r \rho^{\text{WS}}(r) = 1$. For the following study we will employ the values obtained in \cite{DeVries:1987atn},  $R_{\text{Pb}} = 6.624$~fm and $d=  0.549$~fm which have been also  used in \cite{Cepila:2020uxc}. As an alternative to the IPSat model, we consider the scenario where the dipole no longer scatters on individual nucleons, but on the Lorentz contracted nucleus as a whole. Within high energy factorization, the exchanged $t$-channel gluons have delta-like support and it is therefore natural to assume scattering on a Lorentz contracted large nucleus, which gives rise to $A^{1/3}$ scaling of the saturation scale. To include such a scaling within the  BGK model we use 
\begin{align}
  \label{eq:model2}
  \sigma_{q\bar{q}}^{A^{\frac{1}{3}}}(x, r) & = \sigma_0^{\text{Pb}}  \left[ 1 -  \exp \left(-\frac{r^2 \pi^2\alpha_s(\mu_r^2) A^{\frac{1}{3}}xg(x, \mu_r^2) }{3 \sigma_0^{\text{BGK}} }  \right)\right].
\end{align}
Based on simple  geometric considerations, see also \cite{Peredo:2023oym}, one expects the parameter $\sigma_0^{\text{Pb}}$ to scale approximately as  $\sim A^{2/3}$, see also the discussion in \cite{Deganutti:2023qct}. In practice, such a simple rescaling yields still a considerable off-set in the overall normalization. We therefore determine  $\sigma_0^{\text{Pb}}$ from a fit to combined ALICE-CMS data set as  
$ \sigma_0^{\text{Pb}} = (1.587 \pm 0.008)$~barn. Comparing to simple scaling we find that $\sigma_0^{\text{Pb}}/( \sigma_0 \cdot 208^{2/3}) \simeq 1.97$, while the obtained value is approximately $0.63$ times the maximal value of $\sigma_{q\bar{q}}^{\text{IPSat}}(x, r \to \infty)$.

\begin{figure}[t]
  \centering
  \parbox{.49\textwidth}{
     \includegraphics[width=.49\textwidth]{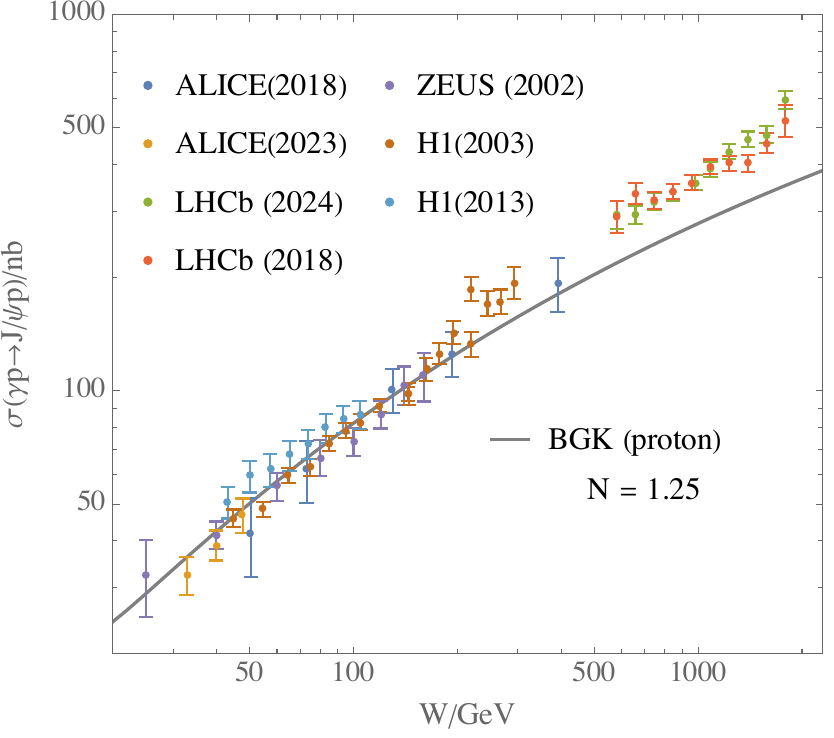}
     }$\,$
 \parbox{.49\textwidth}{
     \includegraphics[width=.49\textwidth]{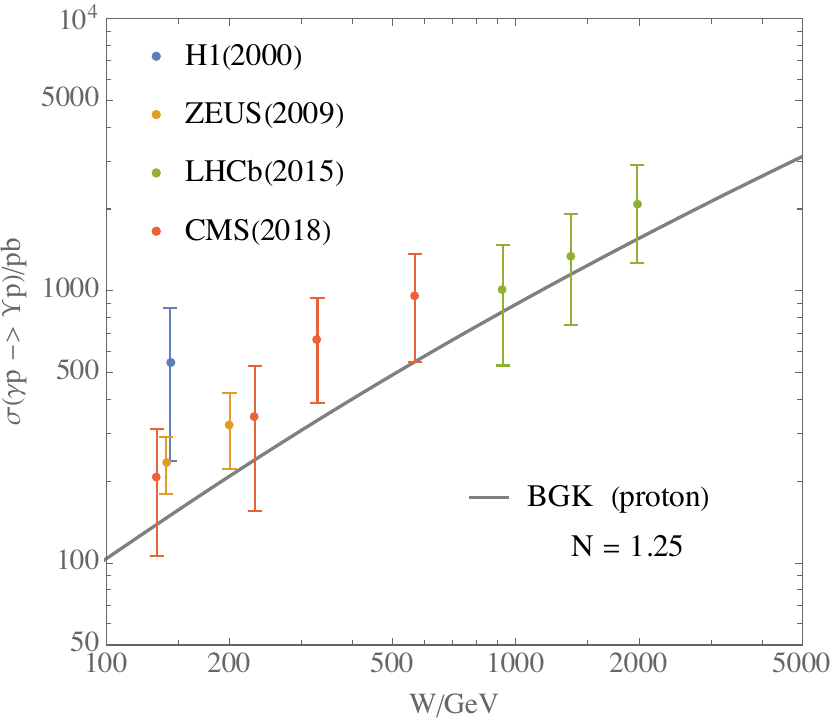}
     }
  \caption{Energy dependence of the total cross-sections for exclusive photoproduction of $J/\Psi$ (left) and  $\Upsilon(1s)$  on a proton, using  the  BGK dipole model.  We further display photo-production data for $J/\psi$  from  ZEUS \cite{ZEUS:2002wfj}
    \cite{Chekanov:2004mw}, H1 \cite{H1:2005dtp,H1:2013okq}
    \cite{Alexa:2013xxa,Aktas:2005xu},
   ALICE \cite{ALICE:2018oyo,ALICE:2023mfc} and LHCb
     \cite{ Aaij:2018arx,LHCb:2024pcz} as well as  $\Upsilon(1s)$ data obtained at   H1
    \cite{H1:2000kis}, ZEUS \cite{ZEUS:2009asc},  LHCb \cite{LHCb:2015wlx} and CMS \cite{CMS:2018bbk}.}
  \label{fig:BGKdata}
\end{figure}

\begin{figure}[th]
  \centering
 \parbox{.49\textwidth}{
     \includegraphics[width=.49\textwidth]{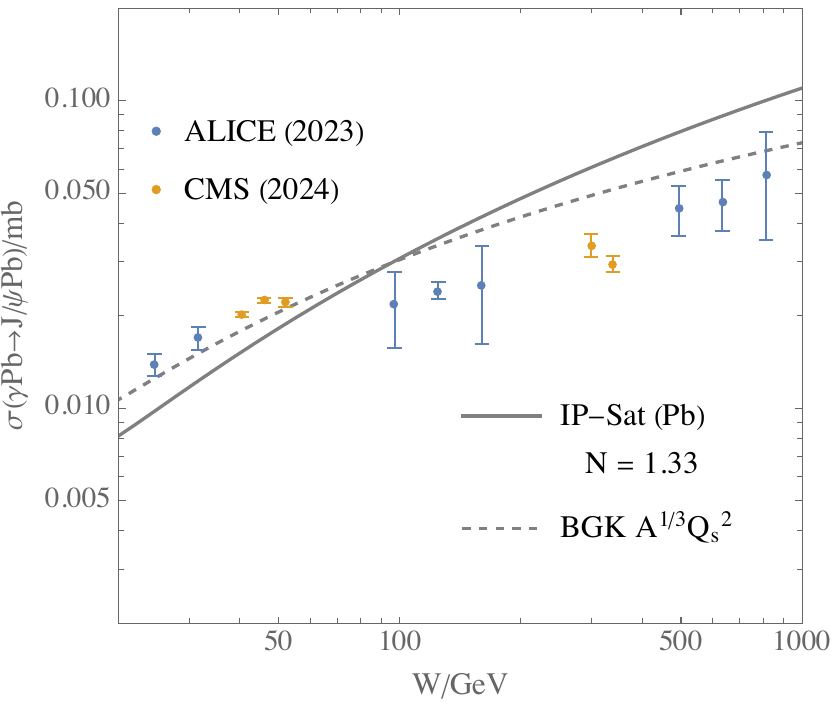}
     }$\,$
 \parbox{.49\textwidth}{
     \includegraphics[width=.49\textwidth]{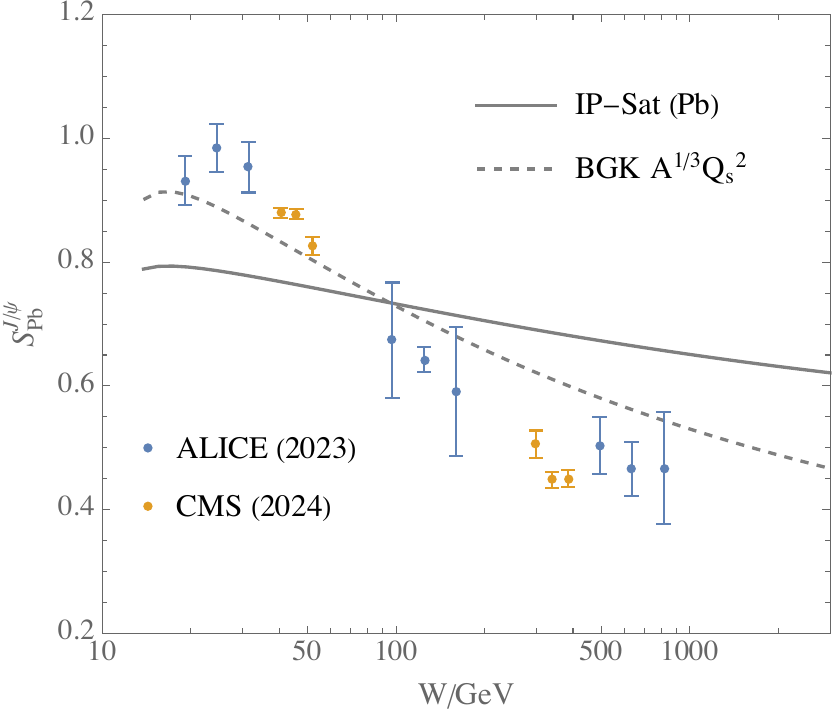}
     }
  \caption{Energy dependence of the  $J/\psi$ photoproduction cross-section on lead (left) and the nuclear modification factor (right) obtained from the dipole models Eq.~\eqref{eq:dsigWS} and Eq.~\eqref{eq:model2}. We further show ALICE \cite{ALICE:2023jgu} and CMS \cite{CMS:2023snh} data.}
  \label{fig:BGKnuc}
\end{figure}

 The energy dependence predicted by these dipoles  models, including a comparison to data is shown in Fig.~\ref{fig:BGKdata} and Fig.~\ref{fig:BGKnuc}. 
Since we are mainly interested in the energy dependence, we allow for adjustments in the overall normalization, through introducing a parameter $N$ through $\Im\text{m}\mathcal{A} \to N \cdot \Im\text{m}\mathcal{A}$, which we adjust through a fit to $J/\psi$ data for $W<500$~GeV. While we find  a very good description of the energy dependence of the $J/\psi$ photoproduction cross-section on a proton, Fig.~\ref{fig:BGKdata}, for  $W<500$~GeV, the model undershoots proton data for $W>500$~GeV. This seems to indicate that the size of saturation effects obtained from the fit to HERA data are too strong. Since we will use in the following the model to create initial conditions at $x = 0.01$, this does not provide a problem to us. For the $\Upsilon(1s)$ photoproduction cross-section we use the normalization determined from the $J/\psi$ fit. We observe that both  energy dependence  and normalization is well described in this case. As a side note we mention that our own  earlier attempts used instead of the BGK model the GBW model; while the energy dependence is still well described for the $\Upsilon$, one finds a considerable offset in normalization, if the latter is fixed using $J/\psi$ data. 
Comparison of the photo-nuclear cross-section with data is shown in Fig.~\ref{fig:BGKnuc}. While the IPSat model provides a good estimate of the overall normalization of the cross-section, which for the $A^{\frac{1}{3}}$ scaled BGK model has been fitted to data, its  energy dependence  is in general too strong and does not agree with data. Even with the proton dipole cross-section undershooting data, the IP-Sat model  -- which has been obtained from the proton dipole cross-section --  overshoots data. The $A^{1/3}$ scaled BGK model provides on the other hand a good description of the energy dependence, in particular if one ignores the off-set with high energy CMS data. Even though the observation is merely made on the levels of saturation models, our results seems to indicate that assuming scattering of the dipole on individual nucleons is no longer justified.  Treating the lead nucleus as a scaled-up proton provides on the other hand a more accurate description of the energy dependence.   We further present results for the nuclear suppression factor which is defined as  $S_{\text{Pb}} = \sqrt{\sigma_{\gamma \text{Pb}}/\sigma_{\gamma \text{Pb}}^{\text{IA}}}$. Following \cite{Guzey:2013xba,ALICE:2023jgu}  the cross-section in the impulse approximation  $\sigma_{\gamma \text{Pb}}^{\text{IA}}$ is obtained as 
\begin{align}
  \label{eq:impulse_approx}
  \sigma_{\gamma \text{Pb}}^{\text{IA}} & = \frac{d\sigma_{\gamma p \to J/\psi p}}{dt}\bigg|_{t=0} \cdot \int_{|t|_{\text{min}}}^{\infty} d|t| \, |F(t)|^2, 
& F(t) & = 208\cdot \int d^3 r e^{i \vec{q}\cdot \vec{r}} \rho^{\text{WS}}(r) , 
\end{align}
with $t = -\vec{q}^2$ and $t_{\text{min}} \simeq 0$ for LHC kinematics. As expected, the IP-Sat model does not describe the energy dependence of the nuclear modification factor. The $A^{\frac{1}{3}}$ scaled BGK model provides on the other hand a good description of data, even though it  overshoots high energy CMS data points.

\section{Energy dependence through BFKL}
\label{sec:bfkl}

In the following  the above dipole models are used to generate initial conditions for BFKL evolution. 

\subsection{BFKL evolution of dipole cross-sections}
\label{sec:bfkl_setup}

Our starting point is the following factorization of the elastic scattering amplitude in the forward limit, $t=0$,  
\begin{align}
  \label{eq:ImA_BFKL}
  \Im\text{m} \mathcal{A}_T^{\gamma p \to V p}(x, 0) & = \int d k^2  \int d q^2 \Phi^{\gamma V}(k)  \mathcal{F}^{\text{DIS}}_{\text{BFKL}}\left(\frac{x_0}{x}, k, q \right)    G(x_0, q),
\end{align}
where $ \mathcal{F}^{\text{DIS}}_{\text{BFKL}}$ denotes the NLO BFKL Green's function in the DIS scheme and 
\begin{align}
  \label{eq:impact_define}
  \Phi^{\gamma V}(k, m_f) & = \frac{4\pi^2}{N_c } \int d^2 \rt  \int_0^1 \frac{dz}{4 \pi}  \frac{1-e^{i  \kt \cdot \rt }}{\kt^2} \left(\Psi_{V}^{*}\Psi_T\right)(r,z),
\end{align}
provides the impact factor for the photon vector meson transition. $G(x_0, q)$ denotes on the other hand the unintegrated gluon distribution at $x_0 = 0.01$. The expression can be constructed as follows: at leading order, one has the following relation between dipole cross-section and unintegrated gluon distribution $ G(x, k)$, 
\begin{align}
  \label{eq:sigqqb_fromugd}
   \sigma_{q\bar{q}}(x, r) & = \frac{4 \pi^2 \alpha_s}{N_c} \int \frac{d^2 \kt}{\pi}  \frac{1-e^{i  \kt \cdot \rt }}{\kt^2} G(x, k), 
& k &= |\kt|,
\end{align}
To evolve $ G(x_0, k)$ to low $x$, we use a solution of the  NLO BFKL Green's function elaborated in  \cite{Hentschinski:2012kr,Hentschinski:2013id, Chachamis:2015ona}
\begin{align}
  \label{eq:Gx_evolved}
  G(x, k) & = \int_0^\infty \frac{d q^2}{q^2} \mathcal{F}^{\text{DIS}}_{\text{BFKL}}\left(\frac{x_0}{x}, k, q \right) G(x_0, q).
\end{align}
The unintegrated gluon distribution at  $x_0$ can be finally determined from the dipole cross-section   using\footnote{See \cite{Luszczak:2022fkf} for a compact summary of the procedure},
\begin{align}
  \label{eq:Gxk}
  G(x_0, k) & = \frac{N_c  k^2}{8 \pi^2 \alpha_s} \int_0^\infty dr \, r J_0(r k) \left[\sigma_0 - \sigma_{q\bar{q}}(x_0, r) \right],
& \sigma_0 & = \lim_{r \to \infty} \sigma_{q\bar{q}}(x_0, r).
\end{align}
Eq.~\eqref{eq:Gx_evolved} describes correctly the low $x$ evolution of the unintegrated gluon distribution if high density corrections are suppressed by powers of the strong coupling $\alpha_s$, {\it i.e. } if the gluon distribution does not saturate the bound $G \sim 1/\alpha_s$. This is the central hypothesis of this section: we assume that such corrections are irrelevant and we  explore to which extend this leads to tension with data. We further note that both Eq.~\eqref{eq:sigqqb_fromugd},\eqref{eq:Gxk} depend on  $\alpha_s$, while the dependence cancels, if both expressions are combined. In general it is not clear whether both strong couplings are  to be evaluated at the same scale, which  gives rise to a normalization ambiguity, see also the discussion in \cite{Luszczak:2022fkf}.

To solve NLO BFKL evolution, we follow closely  \cite{Hentschinski:2012kr,Hentschinski:2013id}. In transverse Mellin space we have
\begin{align}
  \label{eq:Greens1}
   \mathcal{F}^{\text{DIS}}_{\text{BFKL}}\left(\frac{x_0}{x}, k, q \right)  & =\frac{1}{k^2}\int \limits_{\frac{1}{2} - i\infty}^{\frac{1}{2} + i \infty}\frac{d\gamma}{2 \pi i }  \left({k^2}\right)^{\gamma-1} \, \hat{f}\left(\frac{x_0}{x}, \frac{M^2}{Q_s^2(x_0)}, \frac{\overline{M}^2}{M^2}, \gamma \right) \, \, \left({q^2} \right)^{-\gamma},
\end{align}
with $\hat{f}$ an operator in $\gamma$ space and defined as
\begin{align}
  \label{eq:operator}
  \hat{f}  = 
 \left(\frac{1}{x}\right)^{\chi\left(\gamma,  \mu, M \right)} \,\cdot 
  \Bigg\{1    - \frac{\asb^2\beta_0   }{8 N_c} \log{\left(\frac{x_0}{x}\right)} 
  \Bigg[ \overleftarrow{\partial}_\gamma \chi_0(\gamma) - \chi_0(\gamma) \overrightarrow{\partial}_\gamma - 2 \chi_0(\gamma)\ln M^2\Bigg]\Bigg\}\;, 
\end{align}
see also the related discussion in \cite{Chachamis:2015ona,Bautista:2016xnp,SabioVera:2006cza,SabioVera:2007ndx}. Here $\asb = \alpha_s(\mu)N_c/\pi$, while $\mu$ denotes the renormalization scale. $M$ is on the other hand a parameter which has been introduced to identify the central value of the renormalization scale of the problem; it cancels  to NLO accuracy. In the following, we first determine a suitable choice for $M$ and then vary $\mu$ in the range $M/2, 2M$ to estimate the uncertainty associated with our prediction. The determination of $M$ will be discussed below. $\chi(\gamma, \mu, M)$ is the next-to-leading
logarithmic (NLL) BFKL kernel after collinear improvements; in
addition large terms proportional to the first coefficient of the QCD
beta function, $\beta_0 =  11 N_c/3 - 2 n_f /3$  are resummed
through employing a Brodsky-Lepage-Mackenzie (BLM) optimal scale
setting scheme ~\cite{Brodsky:1982gc}. For those details we follow very closely the treatment in \cite{Hentschinski:2012kr,Hentschinski:2013id, Bautista:2016xnp}. The NLL kernel with collinear
improvements reads
\begin{align}\label{eq:gluongf}
\chi\left(\gamma, \mu, M
\right) &= 
{\bar\alpha}_s\chi_0\left(\gamma\right)+
{\bar\alpha}_s^2\tilde{\chi}_1\left(\gamma\right)-\frac{1}{2}{\bar\alpha}_s^2
\chi_0^{\prime}\left(\gamma\right)\chi_0\left(\gamma\right)
\notag \\
&\hspace{4cm}
 +
\chi_{RG}({\bar\alpha}_s,\gamma,\tilde{a},\tilde{b})
 - \frac{\bar{\alpha}_s^2 \beta_0}{4 N_c} \chi_0(\gamma)\log  \frac{\mu^2}{M^2} ,
\end{align}
with the leading-order BFKL eigenvalue, 
\begin{align}
  \label{eq:chi0}
  \chi_0(\gamma) & = 2 \psi(1) - \psi(\gamma) - \psi(1-\gamma) \, .
\end{align}
Employing BLM optimal scale setting and the momentum space (MOM)
physical renormalization scheme based on a symmetric triple gluon
vertex with $Y \simeq 2.343907$ and gauge
parameter $\xi =3$ one obtains the following next-to-leading order
BFKL eigenvalue
\begin{align}
\label{eq:chi1NLO}
\tilde{\chi}_1 (\gamma) &= \tilde{\cal S} \chi_0 (\gamma) + \frac{3}{2} \zeta(3)
+ \frac{ \Psi ''(\gamma) + \Psi''(1-\gamma)- \phi(\gamma)-\phi (1-\gamma) }{4} \nonumber \\
&- \frac{\pi^2 \cos{(\pi \gamma)}}{4 \sin^2{(\pi \gamma)}(1-2\gamma)}
\left[3+\left(1+\frac{n_f}{N_c^3}\right) \frac{2+3\gamma(1-\gamma)}{(3-2\gamma)(1+2\gamma)}\right] \nonumber\\
&+\frac{1}{8} \left[\frac{3}{2} (Y-1) \xi
   +\left(1-\frac{Y}{3}\right) \xi ^2+\frac{17 Y}{2}-\frac{\xi ^3}{6}\right] \chi_0 (\gamma), 
\end{align}
where $\tilde{\cal S} =1/3 - \pi^2/12$. We note that the  specific choice on scheme and parameters follows the one of  \cite{Brodsky:2002ka, Hentschinski:2012kr}, see also the discussion in \cite{Celiberto:2022dyf}.  Within this scheme one then replaces $\asb \to \tilde{\alpha}_s(\gamma)$  with
\begin{eqnarray}
\label{eq:BLM}
\tilde{\alpha}_s \left(\asb, \gamma \right) &=&
                                                                   \frac{\asb(\mu)}{ 1 +  \frac{\asb \beta_0}{4 N_c} \left[ 
\frac{1}{2} \chi_0 (\gamma) - \frac{5}{3} +2 \left(1+ \frac{2}{3} Y\right)\right]},
\end{eqnarray}
which yields the effective running coupling constant to be used for the Green's function. 
For the  numerical study we use the NLO QCD running coupling with $\alpha_s(M_Z) = 0.1181$ and $M_Z = 91.1876$~GeV. 
The term which achieves a resummation of collinear enhanced terms in the NLO BFKL kernel reads
\begin{align}
\label{eq:RG}
 \chi_{RG}(\bar{\alpha}_s, \gamma, a, b)
&  =  \,\,\bar{\alpha}_s (1+ a \bar{\alpha}_s) \left(\psi(\gamma) 
- \psi (\gamma-b \bar{\alpha}_s)\right)  - \frac{\bar{\alpha}_s^2}{2}
  \psi'' (1-\gamma)  -  \frac{ b \bar{\alpha}_s^2 \cdot \pi^2}{\sin^2{(\pi
  \gamma)}}
\notag \\
&
+ \frac{1}{2} \sum_{m=0}^\infty \Bigg(\gamma-1-m+b \bar{\alpha}_s
  - \frac{2 \bar{\alpha}_s (1+a \bar{\alpha}_s)}{1-\gamma+m}
\notag \\
& \hspace{4cm}
+ \sqrt{(\gamma-1-m+b \bar{\alpha}_s)^2+ 4 \bar{\alpha}_s (1+a \bar{\alpha}_s)} \Bigg) \, . 
\end{align}
For details on the derivation of this term we refer to the discussion
in   \cite{Hentschinski:2012kr}.
The coefficients
$\tilde{a}, \tilde{b}$ which enter the collinear resummation term
Eq.~\eqref{eq:RG} are obtained as the coefficients of the $1/\gamma$
and $1/\gamma^2$ poles of the NLO eigenvalue. We have
\begin{align}
  \label{eq:aundb}
  \tilde{a} &=  - \frac{13}{36} \frac{n_f}{N_c^3}- \frac{55}{36} + \frac{3 Y - 3}{16}\xi + \frac{3 - Y}{24} \xi^2 - \frac{1}{48}\xi^3 + \frac{17}{16}Y, \notag\\
\tilde{b} &= - \frac{n_f}{6N_c^3}- \frac{11}{12}.
\end{align}
With the elements of  Eq.~\eqref{eq:operator} fixed,  we  determine as a next step the Mellin transform of $\Phi^{\gamma V}$ and $G(x_0, q)$. We have \cite{Bautista:2016xnp}
\begin{align}
  \label{eq:24}
\phi^{\gamma V}(\gamma, m_f) & = \int_0^\infty \frac{dk^2}{k^2} \left(\frac{k^2}{m_f^2} \right)^\gamma  \Phi^{\gamma V}(k, m_f) =  \int_0^1 \frac{dz}{4 \pi}  \tilde{\phi}_{V,T} (\gamma, z, m_f)  \notag \\
   \tilde{\phi}_{V,T}(\gamma, z, m_f)  &= e \hat{e}_f 8 \pi^2  \mathcal{N}_T \frac{\Gamma(\gamma) \Gamma(1-\gamma)}{m_f^2} \left(\frac{m_f^2 \mathcal{R}^2}{8 z(1-z)} \right)^{2-\gamma}
e^{- \frac{m_f^2 \mathcal{R}^2}{8 z(1-z)} } 
e^{\frac{m_f^2 \mathcal{R}^2}{2}} \, ,
 \notag \\
& \hspace{-1.5cm}
\left[
U\left(2-\gamma, 1, \frac{m_f^2 \mathcal{R}^2}{8z(1-z)}\right) 
+
 [z^2 +
  (1-z)^2] \frac{\Gamma(3-\gamma)}{\Gamma(2-\gamma)}
U\left(3-\gamma, 2, \frac{m_f^2 \mathcal{R}^2}{8z(1-z)}\right)
\right]\, ,
\end{align}
where $U(a,b,z)$ is a hypergeometric function of the second kind or
Kummer's function\footnote{With respect to the expression found in \cite{Bautista:2016xnp}, we corrected an incorrect power of $m_f$ in the exponent as well as relative factor of 2 between the first and the second term}. We note that normalization of the above impact factor is inversely proportional to the heavy quark mass; its precise value takes therefore an important role in fixing the normalization of the photoproduction cross-section. While the heavy quark mass is the natural scale of the impact factor describing the photon-vector meson transition, the choice is less clear for the hadronic impact factor. While one might use as a first guess the saturation scale of the dipole model at initial $x_0 = 0.01$, the NLO BFKL fit to HERA data of  \cite{Hentschinski:2012kr,Hentschinski:2013id} suggests to use a  smaller value, of the order of a few hundred MeV. Within the current setup,  we then define for each unintegrated gluon distribution at $x_0 = 0.01$ a characteristic scale $Q_0$ through
\begin{align}
  \label{eq:define_Q0}
 1- \sigma_{q\bar{q}}(x_0, r= 1/Q_0)/\sigma_0 & = e^{-1};
\end{align}
the resulting scale $Q_0$ is then half the saturation scale of a corresponding GBW type model \cite{Golec-Biernat:2017lfv}. Note that the above choice only affects our result in the way we  fix the  scale $M$. We then finally arrive at
\begin{align}
  \label{eq:phi_ugd}
     \phi^{\text{ugd}}(\gamma, Q_0)) & =   \int_0^\infty {dq^2} \left(\frac{q^2}{Q_0^2} \right)^{-\gamma} G^{\text{ugd}}(x_0, q, Q_0).
\end{align}
To determine this quantity, we first determine numerically $G^{\text{ugd}}(x_0)$ from the dipole cross-section and then from the resulting expression $\phi^{\text{ugd}}(\gamma)$.  With these ingredients we have
\begin{align}
  \label{eq:ImA_Mellin}
  \Im\text{m}\mathcal{A}(x,0) & = N \int \limits_{\frac{1}{2} - i\infty}^{\frac{1}{2} + i \infty}\frac{d\gamma}{2 \pi i }
\left(\frac{m_f^2}{Q_0^2} \right)^\gamma   \left(\frac{x_0}{x}\right)^{\chi\left(\gamma, \mu, M \right)}  
\phi^{\gamma V}(\gamma, m_f) \phi^{\text{ugd}}\left(\gamma, Q_0\right) 
 \notag \\
& 
\hspace{0cm}
\cdot   \Bigg\{1    - \frac{\tilde{\alpha}_s^2(\asb, \gamma)\beta_0  \chi_0 \left(\gamma\right) }{8 N_c} \log{\left(\frac{x_0}{x}\right)} 
\cdot
  \Bigg[ \frac{d}{d\gamma} \ln \frac{\phi^{\gamma V}(\gamma, m_f)}{\phi^{\text{ugd}}(\gamma, Q_0)} - 2 \ln \frac{ M^2}{m_f Q_0}\Bigg]\Bigg\}\;,
\end{align}
where $N$ is a factor that collects the various sources of normalization uncertainty. While $N = \mathcal{O}(1)$, its precise value will be fixed by a comparison to data. We finally set  $M^2 =k \cdot  m_f \cdot Q_0$ with $k$ a parameter which we vary. While $k = 1$ is the natural choice, we will explore also other values in the following; recall that the value of $M$ sets the central value for the renormalization scale $\mu$, which we vary in addition in the range  $[M/2, 2 M]$.

\subsection{Instability of the NLO solution}
\label{sec:instability}
The above solution to the NLO BFKL equation possesses an instability for ultra-low values of $x$, see also the discussion in \cite{Hentschinski:2020yfm,Garcia:2019tne}: for sufficiently small values  of $k$, the second term in the second line of Eq.~\eqref{eq:ImA_Mellin} is negative and growing in magnitude with diminishing $x$. Unlike  \cite{Hentschinski:2020yfm,Garcia:2019tne}, which employed directly the results of the HSS fit,  the effect can be avoided in the current case through choosing a sufficiently large value of $k$. Such a choice comes however at an undesired cost: the growth with energy of the   photoproduction cross-section  is too strong and does not agree with data. We only find agreement with data for a negative NLO correction. Since we would like to explore the degree to which BFKL evolution can be adapted to still describe photoproduction data, we allow in the following for values of $k$ which yield such a contribution. At the same time, we demand that the BFKL prediction is still well justified within QCD perturbation theory. We define to this end two criteria, which our solution must fulfill. We  first define 
\begin{align}
  \label{eq:split}
\Im\text{m}\mathcal{A}(x,0) & = \Im\text{m}\mathcal{A}^{(0)}(x,0) + \Im\text{m}\mathcal{A}^{(1)}(x,0),
\end{align}
where
\begin{align}
  \label{eq:split2}
 \Im\text{m}\mathcal{A}^{(0)}(x,0) & = N \int \limits_{\frac{1}{2} - i\infty}^{\frac{1}{2} + i \infty}\frac{d\gamma}{2 \pi i }
\left(\frac{m_f^2}{Q_0^2} \right)^\gamma   \left(\frac{x_0}{x}\right)^{\chi\left(\gamma, \mu, M \right)}  
\phi^{\gamma V}(\gamma, m_f) \phi^{\text{ugd}}\left(\gamma, Q_0\right) ,
 \notag \\
\Im\text{m}\mathcal{A}^{(1)}(x,0) & = N \int \limits_{\frac{1}{2} - i\infty}^{\frac{1}{2} + i \infty}\frac{d\gamma}{2 \pi i }
\left(\frac{m_f^2}{Q_0^2} \right)^\gamma   \left(\frac{x_0}{x}\right)^{\chi\left(\gamma, \mu, M \right)}  
\phi^{\gamma V}(\gamma, m_f) \phi^{\text{ugd}}\left(\gamma, Q_0\right) 
 \notag \\
& 
\hspace{0cm}
\cdot   \Bigg\{ - \frac{\tilde{\alpha}_s^2(\asb, \gamma)\beta_0  \chi_0 \left(\gamma\right) }{8 N_c} \log{\left(\frac{x_0}{x}\right)} 
\cdot
  \Bigg[ \frac{d}{d\gamma} \ln \frac{\phi^{\gamma V}(\gamma, m_f)}{\phi^{\text{ugd}}(\gamma, Q_s^2(x_0))} - 2 \ln k \Bigg]\Bigg\}\;,
\end{align}
see also the discussion in \cite{Garcia:2019tne}. 
Given these expressions we can define now 2 criteria which indicate that the choice of $k$ is no longer natural and  stability of the perturbative description is to be lost:
\begin{itemize}
\item  First we require that 
 $| \Im\text{m}\mathcal{A}^{(1)}(x^{\text{min}},0)|/ \Im\text{m}\mathcal{A}^{(0)}(x^{\text{min}},0) < 0.5$ at the lowest observed value of $x$; for the proton $x^{\text{min}}_p \simeq 3.0 \cdot 10^{-6}$, whereas for lead one currently has $x^{\text{min}}_{Pb} \simeq 1.45 \cdot 10^{-5}$.
\item   A second criteria is that we require a positive slope of the scattering amplitude $\lambda > 0$, i.e. $\Im\text{m}\mathcal{A}(x^{\text{min}}, 0)$ must still  grow with energy.
\end{itemize}
For $J/\psi$ photo-production, these criteria allow to  push the BFKL prediction to the limits of its applicability. Photoproduction of the $\Upsilon$ provides on the other hand an important cross-check since one expects a stable perturbative prediction in this case.

\section{Results and discussion}
\label{sec:num}

\begin{figure}[th]
  \centering
  \includegraphics[width=.49\textwidth]{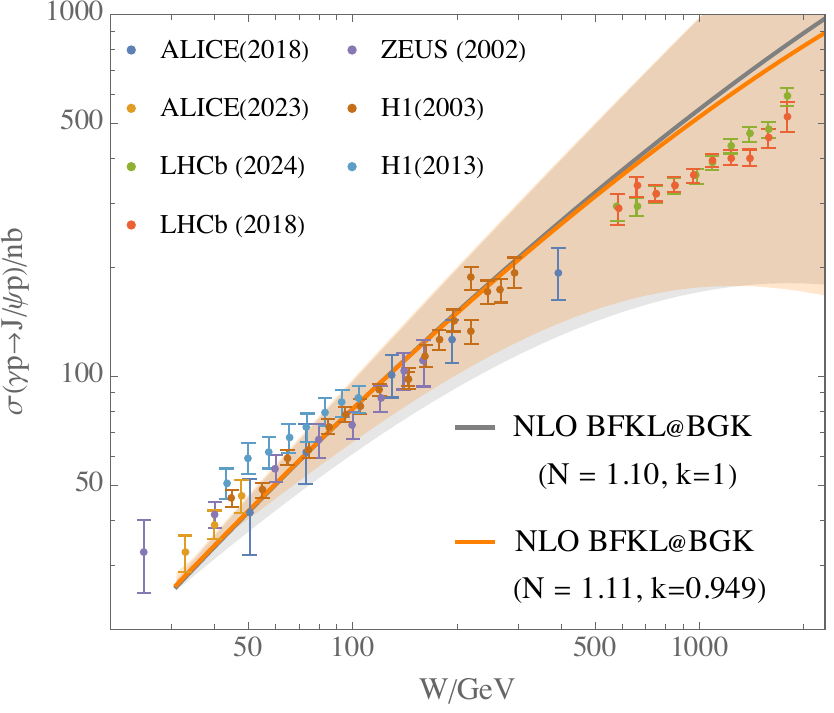} 
$\,$
\includegraphics[width=.49\textwidth]{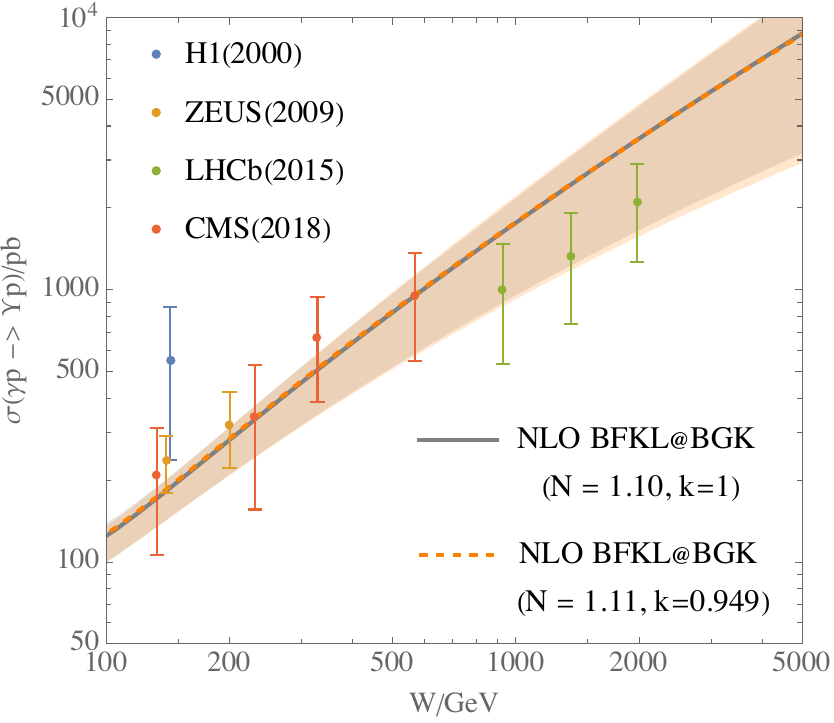}  
  \caption{Energy dependence of the $J/\psi$ (left) and $\Upsilon$ (right) photoproduction cross-section  as obtained from acting with the  NLO BFKL Green's function on the  BGK  dipole cross-section at $x_0= 0.01$.  We displayed the same  data sets as in Fig.~\ref{fig:BGKdata}}
 \label{fig:bfklprot}
\end{figure}

As a first step, we provide predictions for the energy dependence of the photoproduction on the proton, both for $J/\psi$ and $\Upsilon$ vector mesons. Condition Eq.~\eqref{eq:define_Q0} yields in this case $Q_0^{\text{prot.}} = 0.256$~GeV, which is numerically very close to the scale of the impact factor found in \cite{Hentschinski:2012kr,Hentschinski:2013id}. Our results are shown in Fig.~\ref{fig:bfklprot}.
We would like to stress that even for the proton case, the ability of NLO BFKL evolution to describe the energy dependence of photoproduction is a non-trivial observation. The initial conditions for the unintegrated gluon distribution where not fitted to collider data, employing NLO BFKL evolution as the underlying evolution equation. Instead, the initial conditions have been obtained from   the BGK dipole model, which itself has been fitted to inclusive DIS data. The fact that NLO BFKL evolution provides a good description of the energy dependence of both the $J/\psi$ and $\Upsilon$ photoproduction cross-section is yet another confirmation that NLO BFKL is in general terms a useful framework to predict the energy dependence of perturbative QCD cross-sections. We both study the scenario of $k=1$ as well as of minimal $k = 0.949$. The deviation of the overall normalization constant is very close to one and has been fixed through a fit to $J/\psi$ data at $W<500$~GeV. The central values of the BFKL prediction overshoot in both cases data points at $W>500$~GeV. An improved description  with a smaller value of $k$, would however violate our stability conditions, formulated in Sec.~\ref{sec:instability}. 
The depicted uncertainty band is obtained due to a variation of the renormalization scale in the range $\mu \in [M/2, 2\cdot M]$. For $W>500$~GeV, the  uncertainty associated with our prediction of  the $J/\psi$ photoproduction cross-section is  already considerable. While this is unpleasant, such a behavior was to be expected, based on the results of \cite{Garcia:2019tne, Hentschinski:2020yfm}. As argued there, the BFKL prediction leaves certain space for non-linear QCD evolution -- which in some cases can provide an improved description -- but the uncertainties are too large to provide definite evidence. Uncertainties  are considerably reduced for  $\Upsilon$ photoproduction. While our predictions agree within uncertainties with experimental results, our central predictions overshoots data. This is contrast to the result of \cite{Penttala:2024hvp}, where BFKL was found to provide a good description of the energy dependence. Since  \cite{Penttala:2024hvp} uses a $t$-dependent solution to the BFKL equation, this might indicate a stronger growth of the diffractive slope parameter $B_D$ than provided by the parametrization of HERA data Eq.~\eqref{eq:18}. 
Last but not least we would like to comment that one may also try to obtain the dipole distribution function using the conceptually simpler Golec-Biernat W\"usthof (GBW) dipole model. While this model has the advantage that analytic expressions can be obtained for the Mellin transform of the unintegrated gluon distribution, it lacks corrections due to DGLAP evolution at the initial $x_0$, which gives rise to a certain off-set in normalization between the $J/\psi$ and $\Upsilon$ photoproduction  cross-section. While not central to the objective of this paper, we believe that this is an interesting observation since it indicates that DGLAP type corrections are also needed for impact factors.  

\begin{figure}[th]
  \centering
 \centering
  \includegraphics[width=.49\textwidth]{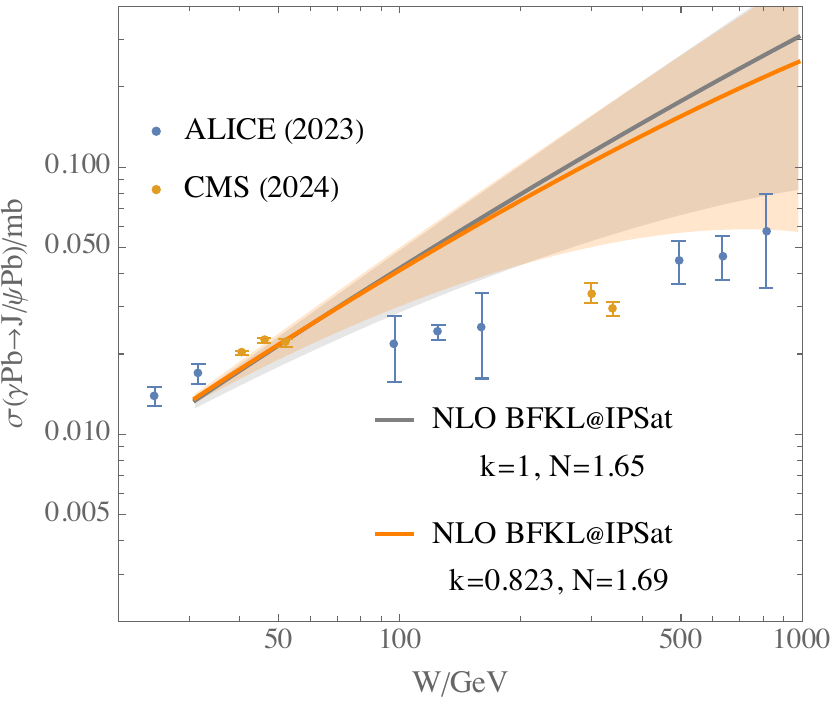} 
$\,$
\includegraphics[width=.49\textwidth]{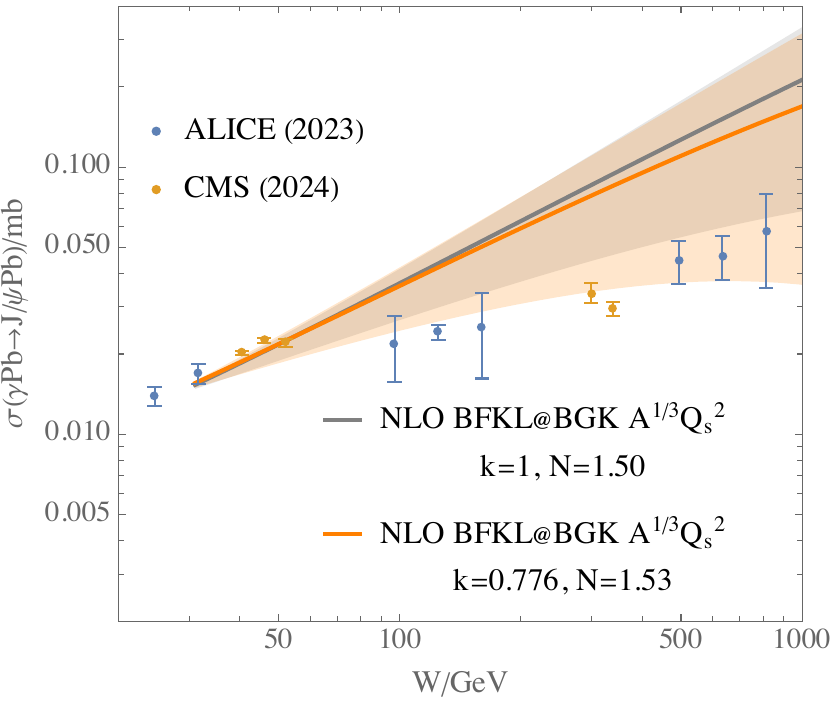}  
  \caption{Energy dependence of the $J/\psi$   photoproduction cross-section  as obtained from acting with the  NLO BFKL Green's function on the  IP-Sat (left) and $A^{\frac{1}{3}}$ scaled BGK  dipole cross-section at $x_0= 0.01$.  We displayed the same  data sets as in Fig.~\ref{fig:BGKnuc}}
  \label{fig:BFKLjpsi_lead}
\end{figure}

BFKL evolution for  $J/\psi$ photoproduction on lead is shown in Fig.~\ref{fig:BFKLjpsi_lead}, including a comparison to data. 
For the IP-Sat model we find $Q_0^{\text{IPs}} = 0.344$~GeV, which is only marginally larger than the proton results. For the $A^{\frac{1}{3}}$ scaled proton saturation scale, we have instead $Q_0^{A^{\frac{1}{3}}}= 0.622$~GeV. As for the proton, we show predictions both for the case $k=1$ and for the minimal $k$ value, which is allowed by  the stability requirements of Sec.~\ref{sec:instability}. While  the higher  values of $Q_0$ allow to extend the prescription down to lower values of $k$, the impact on the energy dependence is rather small. For the determination of the normalization, we exclude data points with $W>280$~GeV. The resulting adjustments in normalization are larger than in the case of the proton. We note that not fixing the normalization leads to an improved description of values at $W>280$~GeV, while one clearly misses the low energy data points. Since our main goal is to study predictions for the  energy dependence, fixing the low energy normalization  appears to be the right choice here. 
\begin{figure}[th]
  \centering
 \centering
  \includegraphics[width=.49\textwidth]{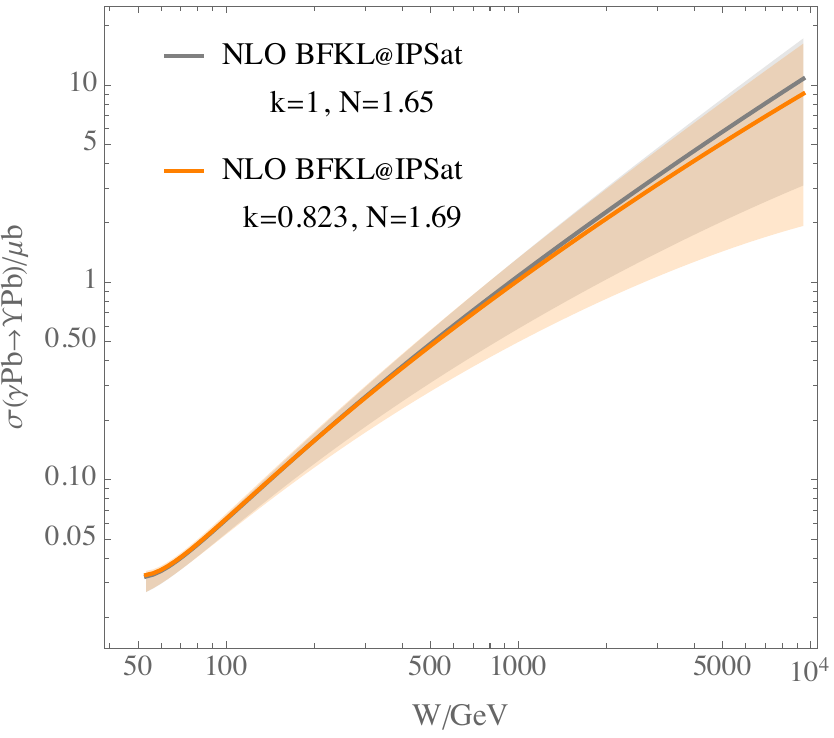} 
$\,$
\includegraphics[width=.49\textwidth]{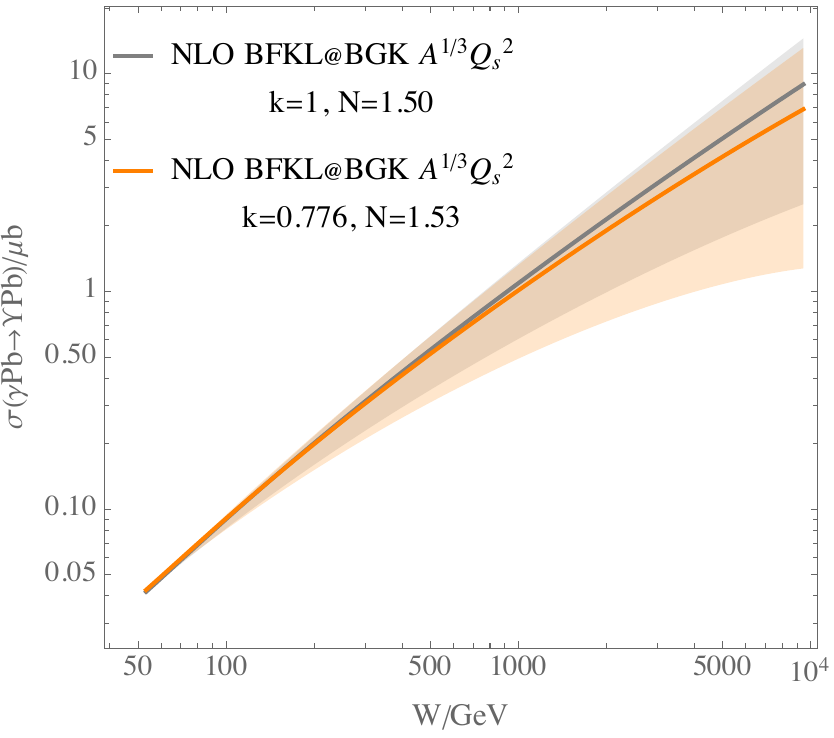}  
  \caption{Energy dependence of the $\Upsilon$   photoproduction cross-section  as obtained from acting with the  NLO BFKL Green's function on the  IP-Sat (left) and $A^{\frac{1}{3}}$ scaled BGK  dipole cross-section at $x_0= 0.01$.}
  \label{fig:BFKLupsi_lead}
\end{figure}
If initial conditions for BFKL evolution are generated through the IPSat model, we find  strong tension between the obtained energy dependence and data. Assuming that the IP-Sat model provides the correct description of the dipole cross-section at values of $x> 0.01$, we can exclude almost with certainty that the energy dependence of photonuclear $J/\psi$ production is described by low density QCD low $x$ evolution. While the $A^{1/3}$ scaled BGK model also provides a BFKL prediction which clearly overshoots data, the situation is less clear in that case. In particular for $k =0.776$, the majority of data points is still described within the large uncertainty of the BFKL prediction. At the same time, one needs to keep in mind that the description is based on a significantly increased value of $Q_0$, which requires strong correlations between individual nucleons in the lead nucleus. 

In Fig.~\ref{fig:BFKLupsi_lead} we finally provide predictions for $\Upsilon$ photonuclear production. As in the proton case, uncertainties are reduced for the $\Upsilon$ due to the higher hard scale. Similar to the $J/\psi$ case we find that BFKL based on IPSat initial conditions grows stronger with energy than the $A^{1/3}$ scaled BGK saturation model. 

\begin{figure}[th]
  \centering
   \includegraphics[width=.49\textwidth]{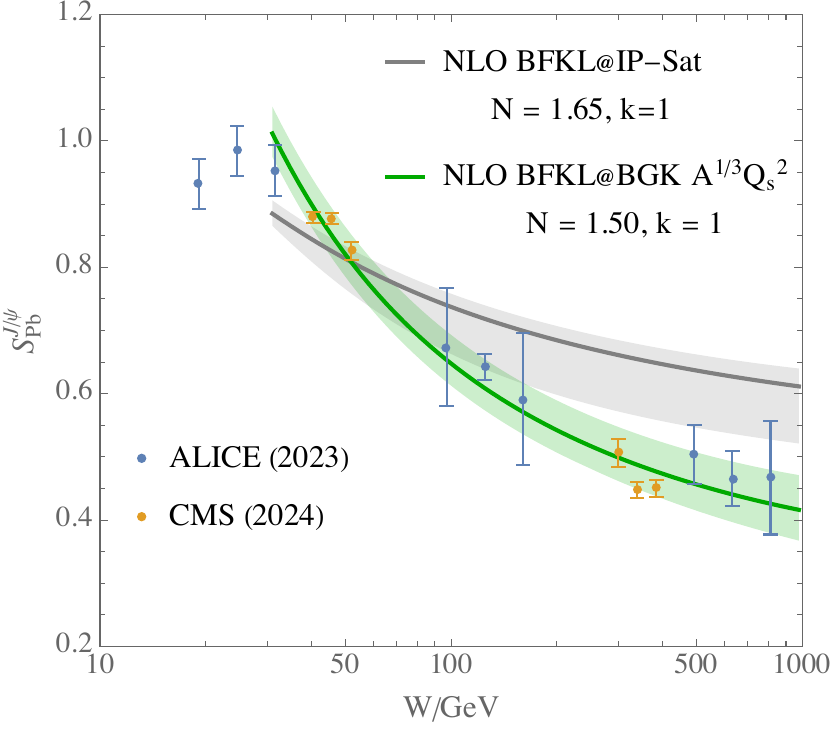} 
$\,$
\includegraphics[width=.49\textwidth]{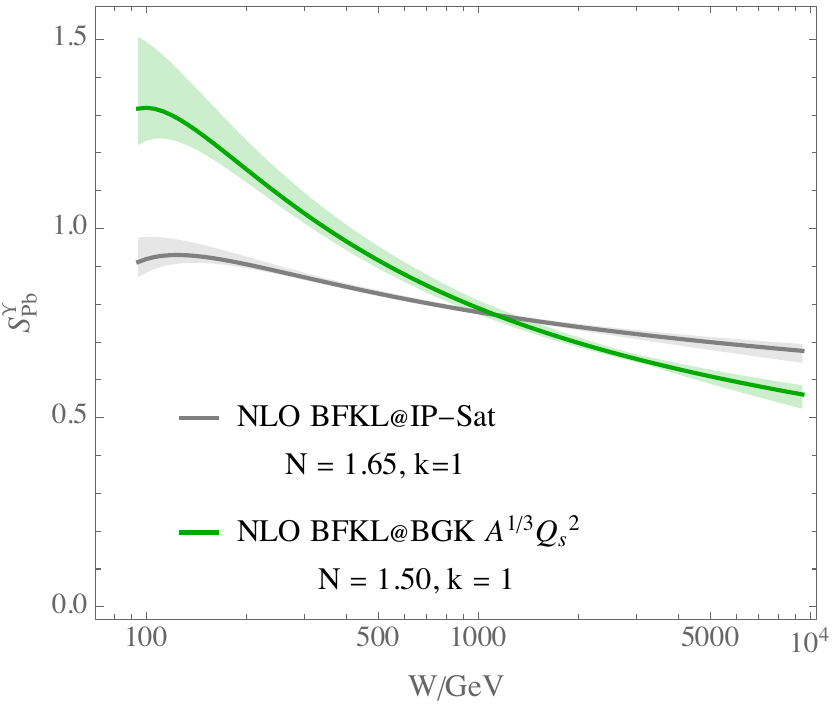}  
  \caption{Nuclear modification factor for $J/\psi$ (left) and $\Upsilon$ (right).  We further show ALICE \cite{ALICE:2023jgu} and CMS \cite{CMS:2023snh} data.}
  \label{fig:nucmod_bfkl}
\end{figure}

In Fig.~\ref{fig:nucmod_bfkl} we finally provide the nuclear modification factor for both $J/\psi$ and $\Upsilon$ photonuclear production. 
As for saturation models we determine the nuclear modification factor as $S_{\gamma\text{Pb}} = \sqrt{\sigma_{\gamma\text{Pb}}/\sigma^{\text{IA}}_{\gamma\text{Pb}}}$, where $\sigma^{\text{IA}}_{\gamma\text{Pb}}$ is determined from Eq.~\eqref{eq:impulse_approx}, using now the proton cross-section with BFKL evolution. As one might expect, the theoretical uncertainties associated with BFKL evolution  reduce significantly for the nuclear modification factor $S^{J/\psi}_{\gamma\text{Pb}}$. While the energy dependence predicted by the NLO BFKL evolution with  IPSat model initial conditions is within uncertainties in agreement with ALICE data,  it clearly overshoots CMS data points at high $W$. The description based on  NLO BFKL evolution based on  $A^{\frac{1}{3}}$ scaled BGK model describes the energy dependence of the nuclear modification factor very well, over the entire range in energies. For photonuclear production of $\Upsilon(1s)$,  the nuclear modification factor is within our framework significantly reduced, albeit the deviation from one is still sizable at highest values of $W$.

\section{Conclusions}
\label{sec:concl}

We explored to which extend NLO BFKL evolution is able to describe the energy dependence of photoproduction cross-sections of vector mesons $J/\psi$ and $\Upsilon(1s)$  on a proton and on a lead nucleus. Unlike previous studies, e.g. \cite{Penttala:2024hvp}, which provided a solution of the BFKL equation for non-zero momentum transfer $t\neq 0$, but were restricted to leading order in the perturbative expansion, our solution is restricted to $t=0$, while the complete NLO corrections have been included, together with a resummation of collinear enhanced terms in the NLO BFKL Green's function. As a first result we find that we need to push the NLO BFKL description to the limits of its applicability  to describe the observed energy dependence. To guarantee perturbative stability, we defined two criteria, i.e. a limit on the size of a perturbative correction to the Green's function as well as a positive slope at smallest observed $x$ values. With these criteria we find that for $J/\psi$ photoproduction, central values of the BFKL description  overshoot both proton and lead data. While proton data agree with the BFKL description within the sizable theory uncertainties,  lead data only agree with the BFKL prediction within uncertainties, if the $A^{\frac{1}{3}}$ scaled BGK model is used to generate initial conditions for  BFKL evolution. The photoproduction cross-section of the  $\Upsilon$  on the proton is on the other hand very well described. 

A natural question to ask is whether an improved description would have been obtained, if a solution to the NLO BFKL equation with $t \neq 0$ would have been used, similar to the study presented in \cite{Penttala:2024hvp}. We believe that this  is not the case, since the observed energy dependence of the  diffractive slope $B_D$ is in general weak and should not lead to such a drastic change. Instead it seems more likely that a more suitable choice of initial conditions could provide an improved description of data. 
On the other hand, \cite{Penttala:2024hvp} find a strong dependence on the impact parameter in their solution to the BFKL evolution equation in the case of the proton, which is somehow in contradiction to this line of argument.   While we are currently not able comment on the details of the solution of \cite{Penttala:2024hvp}, we believe that it would be a worthwhile exercise to compare in the future both solutions as well as to explore possibilities to arrive at   NLO BFKL predictions with $t$-dependence, see \cite{Chachamis:2022jis} for a framework which might allow for progress in this direction.  For the time being, a  particularly interesting observable is provided by the nuclear modification factor: we find that the sizable uncertainties of the BFKL description reduces significantly in this case and we find an excellent description of data for initial conditions created by the $A^{\frac{1}{3}}$ scaled BGK model, whereas BFKL evolution based on the IPSat model clearly overshoots data.

Our result indicate the following: at first  there are strong indications that the nuclear gluon density per transverse area scales as $A^{\frac{1}{3}}$. The $A^{\frac{1}{3}}$  scaled dipole model provides itself  a good description of the photonuclear cross-section as well as of the  nuclear modification factor. At the same time, the $A^{\frac{1}{3}}$  scaled dipole model provides adequate initial conditions for NLO BFKL evolution. This is  particularly remarkable since  descriptions based on the IPSat model fails. Our result for the nuclear modification factor seems furthermore to indicate that a description of the energy dependence does currently not require the inclusion of non-linear terms into  low $x$ evolution. Nevertheless one should keep in mind that the NLO BFKL description is only possible due to the negative perturbative correction to the NLO BFKL Green's function. Choosing the renormalization scale  in a way such that this term is very small or positive, a description of the energy dependence would be not possible. In this sense, our result cannot exclude completely the need of a non-linear terms in the evolution (which arguably would have a similar effect). However our results seem to imply that the size of non-linear terms should be of similar magnitude for both photoproduction on the proton and a lead nucleus. 

For future studies it would be interesting to generate initial conditions for NLO BFKL evolution directly from nuclear PDFs. While this is challenging, it would probably reduce further the uncertainty in fixing initial conditions. A future measurement of the $\Upsilon$ production at lead targets would be of great use to further verify the current approach. While one expects non-linear dynamics to further weaken in this case, different initial conditions to NLO BFKL evolution provide significantly different results for the nuclear modification factor in the case of the $\Upsilon$; its measurement would therefore clearly contribute to clarify the situation further and to fix the correct initial conditions. Finally, measurement of the dependence of the vector meson production cross-section on photon virtuality at the future EIC, would allow determine the lead impact factor from a fit to EIC data and therefore reduce uncertainties.


\section*{Acknowledgments}

We acknowledge support by IANN-QCD network and discussions with Zhoudunming~Tu. MH is grateful to the BNL EIC Theory institute for financial support and to Krzysztof  Kutak and members of the BNL nuclear theory group for useful discussions.  

\bibliographystyle{hunsrt} 

\begin{thebibliography}{99}
\bibitem{Gribov:1984tu}
L.~V.~Gribov, E.~M.~Levin and M.~G.~Ryskin,
Phys. Rept. \textbf{100}, 1-150 (1983)
doi:10.1016/0370-1573(83)90022-4

\bibitem{Morreale:2021pnn}
A.~Morreale and F.~Salazar,
Universe \textbf{7}, no.8, 312 (2021)
doi:10.3390/universe7080312
[arXiv:2108.08254 [hep-ph]].

\bibitem{Hentschinski:2022xnd}
M.~Hentschinski, C.~Royon, M.~A.~Peredo, C.~Baldenegro, A.~Bellora, R.~Boussarie, F.~G.~Celiberto, S.~Cerci, G.~Chachamis and J.~G.~Contreras, \textit{et al.}
Acta Phys. Polon. B \textbf{54}, no.3, 3-A2 (2023)
doi:10.5506/APhysPolB.54.3-A2
[arXiv:2203.08129 [hep-ph]].

\bibitem{Aguilar:2024otb}
A.~C.~Aguilar, A.~Bashir, J.~J.~Cobos-Mart{\'\i}nez, A.~Courtoy, B.~El-Bennich, D.~de Florian, T.~Frederico, V.~P.~Gon{\c{c}}alves, M.~Hentschinski and R.~J.~Hern{\'a}ndez-Pinto, \textit{et al.}
Braz. J. Phys. \textbf{55}, no.4, 145 (2025)
doi:10.1007/s13538-025-01778-x
[arXiv:2409.18407 [nucl-ex]].

\bibitem{Kuraev:1977fs}
E.~A.~Kuraev, L.~N.~Lipatov and V.~S.~Fadin,
Sov. Phys. JETP \textbf{45}, 199-204 (1977)

\bibitem{Kuraev:1976ge}
E.~A.~Kuraev, L.~N.~Lipatov and V.~S.~Fadin,
Sov. Phys. JETP \textbf{44}, no.3, 443-451 (1976)

\bibitem{Balitsky:1978ic}
I.~I.~Balitsky and L.~N.~Lipatov,
Sov. J. Nucl. Phys. \textbf{28}, 822-829 (1978)

\bibitem{Ferreiro:2001qy}
E.~Ferreiro, E.~Iancu, A.~Leonidov and L.~McLerran,
Nucl. Phys. A \textbf{703} (2002), 489-538
doi:10.1016/S0375-9474(01)01329-X
[arXiv:hep-ph/0109115 [hep-ph]].

\bibitem{Iancu:2001ad}
E.~Iancu, A.~Leonidov and L.~D.~McLerran,
Phys. Lett. B \textbf{510} (2001), 133-144
doi:10.1016/S0370-2693(01)00524-X
[arXiv:hep-ph/0102009 [hep-ph]].

\bibitem{Iancu:2000hn}
E.~Iancu, A.~Leonidov and L.~D.~McLerran,
Nucl. Phys. A \textbf{692} (2001), 583-645
doi:10.1016/S0375-9474(01)00642-X
[arXiv:hep-ph/0011241 [hep-ph]].

\bibitem{Weigert:2000gi}
H.~Weigert,
Nucl. Phys. A \textbf{703} (2002), 823-860
doi:10.1016/S0375-9474(01)01668-2
[arXiv:hep-ph/0004044 [hep-ph]].

\bibitem{Kovchegov:1999yj}
Y.~V.~Kovchegov,
Phys. Rev. D \textbf{60} (1999), 034008
doi:10.1103/PhysRevD.60.034008
[arXiv:hep-ph/9901281 [hep-ph]].

\bibitem{Jalilian-Marian:1997ubg}
J.~Jalilian-Marian, A.~Kovner and H.~Weigert,
Phys. Rev. D \textbf{59} (1998), 014015
doi:10.1103/PhysRevD.59.014015
[arXiv:hep-ph/9709432 [hep-ph]].

\bibitem{Balitsky:1995ub}
I.~Balitsky,
Nucl. Phys. B \textbf{463} (1996), 99-160
doi:10.1016/0550-3213(95)00638-9
[arXiv:hep-ph/9509348 [hep-ph]].



\bibitem{Bautista:2016xnp}
I.~Bautista, A.~Fernandez Tellez and M.~Hentschinski,
Phys. Rev. D \textbf{94}, no.5, 054002 (2016)
doi:10.1103/PhysRevD.94.054002
[arXiv:1607.05203 [hep-ph]].

\bibitem{Cepila:2018faq}
J.~Cepila, J.~G.~Contreras and M.~Matas,
Phys. Rev. D \textbf{99}, no.5, 051502 (2019)
doi:10.1103/PhysRevD.99.051502
[arXiv:1812.02548 [hep-ph]].


\bibitem{Garcia:2019tne}
A.~Arroyo Garcia, M.~Hentschinski and K.~Kutak,
Phys. Lett. B \textbf{795}, 569-575 (2019)
doi:10.1016/j.physletb.2019.06.061
[arXiv:1904.04394 [hep-ph]].



\bibitem{Krelina:2019gee}
M.~Krelina, V.~P.~Goncalves and J.~Cepila,
Nucl. Phys. A \textbf{989}, 187-200 (2019)
doi:10.1016/j.nuclphysa.2019.06.009
[arXiv:1905.06759 [hep-ph]].

\bibitem{Klein:2019qfb}
S.~R.~Klein and H.~M{\"a}ntysaari,
Nature Rev. Phys. \textbf{1}, no.11, 662-674 (2019)
doi:10.1038/s42254-019-0107-6
[arXiv:1910.10858 [hep-ex]].

\bibitem{Kopeliovich:2020has}
B.~Z.~Kopeliovich, M.~Krelina, J.~Nemchik and I.~K.~Potashnikova,
Phys. Rev. D \textbf{107}, no.5, 054005 (2023)
doi:10.1103/PhysRevD.107.054005
[arXiv:2008.05116 [hep-ph]].

\bibitem{Bendova:2020hbb}
D.~Bendova, J.~Cepila, J.~G.~Contreras and M.~Matas,
Phys. Lett. B \textbf{817}, 136306 (2021)
doi:10.1016/j.physletb.2021.136306
[arXiv:2006.12980 [hep-ph]].


\bibitem{Hentschinski:2020yfm}
M.~Hentschinski and E.~Padr{\'o}n Molina,
Phys. Rev. D \textbf{103}, no.7, 074008 (2021)
doi:10.1103/PhysRevD.103.074008
[arXiv:2011.02640 [hep-ph]].

\bibitem{Jenkovszky:2021sis}
L.~Jenkovszky, V.~Libov and M.~V.~T.~Machado,
Phys. Lett. B \textbf{824}, 136836 (2022)
doi:10.1016/j.physletb.2021.136836
[arXiv:2111.13389 [hep-ph]].

\bibitem{Flett:2021xsl}
C.~A.~Flett,

\bibitem{Mantysaari:2021ryb}
H.~M{\"a}ntysaari and J.~Penttala,
Phys. Lett. B \textbf{823}, 136723 (2021)
doi:10.1016/j.physletb.2021.136723
[arXiv:2104.02349 [hep-ph]].


\bibitem{Mantysaari:2022kdm}
H.~M{\"a}ntysaari and J.~Penttala,
JHEP \textbf{08}, 247 (2022)
doi:10.1007/JHEP08(2022)247
[arXiv:2204.14031 [hep-ph]].

\bibitem{Goncalves:2022ret}
V.~P.~Goncalves, B.~D.~Moreira and L.~Santana,
Phys. Rev. C \textbf{107}, no.5, 055205 (2023)
doi:10.1103/PhysRevC.107.055205
[arXiv:2210.11911 [hep-ph]].

\bibitem{Wang:2022vhr}
X.~Y.~Wang, F.~Zeng and Q.~Wang,
Phys. Rev. D \textbf{105}, no.9, 096033 (2022)
doi:10.1103/PhysRevD.105.096033
[arXiv:2204.07294 [hep-ph]].


\bibitem{Mantysaari:2023xcu}
H.~M{\"a}ntysaari, F.~Salazar and B.~Schenke,
Phys. Rev. D \textbf{109}, no.7, L071504 (2024)
doi:10.1103/PhysRevD.109.L071504
[arXiv:2312.04194 [hep-ph]].


\bibitem{Cepila:2023dxn}
J.~Cepila, J.~G.~Contreras, M.~Matas and A.~Ridzikova,
Phys. Lett. B \textbf{852}, 138613 (2024)
doi:10.1016/j.physletb.2024.138613
[arXiv:2312.11320 [hep-ph]].


\bibitem{Mantysaari:2024zxq}
H.~M{\"a}ntysaari, J.~Penttala, F.~Salazar and B.~Schenke,
Phys. Rev. D \textbf{111}, no.5, 5 (2025)
doi:10.1103/PhysRevD.111.054033
[arXiv:2411.13533 [hep-ph]].

\bibitem{Cepila:2024qge}
J.~Cepila, J.~G.~Contreras, M.~Matas and M.~Vaculciak,
Phys. Rev. D \textbf{111}, no.9, 096015 (2025)
doi:10.1103/PhysRevD.111.096015
[arXiv:2412.08571 [hep-ph]].

\bibitem{Penttala:2024hvp}
J.~Penttala and C.~Royon,
Phys. Lett. B \textbf{864}, 139394 (2025)
doi:10.1016/j.physletb.2025.139394
[arXiv:2411.14815 [hep-ph]].



\bibitem{Nemchik:2025myg}
J.~Nemchik and J.~{\'O}bertov{\'a},
Phys. Rev. D \textbf{112}, no.9, 094005 (2025)
doi:10.1103/hmd1-64yj
[arXiv:2507.01531 [hep-ph]].

\bibitem{Goncalves:2024jlx}
V.~P.~Goncalves, B.~D.~Moreira and L.~Santana,
Eur. Phys. J. C \textbf{84}, no.9, 893 (2024)
doi:10.1140/epjc/s10052-024-13277-5
[arXiv:2404.02746 [hep-ph]].

\bibitem{Cepila:2025rkn}
J.~Cepila, J.~G.~Contreras and M.~Vaculciak,
Phys. Rev. D \textbf{111}, no.5, 056002 (2025)
doi:10.1103/PhysRevD.111.056002
[arXiv:2501.09462 [hep-ph]].

\bibitem{Klein:2020nvu}
S.~Klein, D.~Tapia Takaki, J.~Adam, C.~Aidala, A.~Angerami, B.~Audurier, C.~Bertulani, C.~Bierlich, B.~Blok and J.~D.~Brandenburg, \textit{et al.}
[arXiv:2009.03838 [hep-ph]].

\bibitem{Bylinkin:2022wkm}
A.~Bylinkin, J.~Nystrand and D.~Tapia Takaki,
J. Phys. G \textbf{50} (2023) no.5, 055105
doi:10.1088/1361-6471/acc419
[arXiv:2211.16107 [nucl-ex]].


\bibitem{ALICE:2023fov}
 [ALICE],
ALICE-PUBLIC-2023-001.

\bibitem{daCosta:2025frd}
P.~E.~A.~da Costa, A.~V.~Giannini, V.~P.~Goncalves and B.~D.~Moreira,
Phys. Rev. D \textbf{112} (2025) no.3, 034012
doi:10.1103/2h3t-ww4j
[arXiv:2505.14019 [nucl-ex]].

\bibitem{Amoroso:2022eow}
S.~Amoroso, A.~Apyan, N.~Armesto, R.~D.~Ball, V.~Bertone, C.~Bissolotti, J.~Bluemlein, R.~Boughezal, G.~Bozzi and D.~Britzger, \textit{et al.}
Acta Phys. Polon. B \textbf{53} (2022) no.12, 12-A1
doi:10.5506/APhysPolB.53.12-A1
[arXiv:2203.13923 [hep-ph]].



\bibitem{Frankfurt:2022jns}
L.~Frankfurt, V.~Guzey, A.~Stasto and M.~Strikman,
Rept. Prog. Phys. \textbf{85} (2022) no.12, 126301
doi:10.1088/1361-6633/ac8228
[arXiv:2203.12289 [hep-ph]].

\bibitem{Arleo:2025oos}
F.~Arleo, P.~Caucal, A.~Deshpande, J.~M.~Durham, G.~M.~Innocenti, J.~Jalilian-Marian, A.~Kusina, M.~X.~Liu, Y.~Mehtar-Tani and C.~J.~Na{\"\i}m, \textit{et al.}
[arXiv:2506.17454 [hep-ph]].


\bibitem{ALICE:2023jgu}
S.~Acharya \textit{et al.} [ALICE],
JHEP \textbf{10}, 119 (2023)
doi:10.1007/JHEP10(2023)119
[arXiv:2305.19060 [nucl-ex]].

\bibitem{CMS:2023snh}
A.~Tumasyan \textit{et al.} [CMS],
Phys. Rev. Lett. \textbf{131}, no.26, 262301 (2023)
doi:10.1103/PhysRevLett.131.262301
[arXiv:2303.16984 [nucl-ex]].

\bibitem{Hentschinski:2012kr}
M.~Hentschinski, A.~Sabio Vera and C.~Salas,
Phys. Rev. Lett. \textbf{110}, no.4, 041601 (2013)
doi:10.1103/PhysRevLett.110.041601
[arXiv:1209.1353 [hep-ph]].

\bibitem{Hentschinski:2013id}
M.~Hentschinski, A.~Sabio Vera and C.~Salas,
Phys. Rev. D \textbf{87}, no.7, 076005 (2013)
doi:10.1103/PhysRevD.87.076005
[arXiv:1301.5283 [hep-ph]].

\bibitem{Golec-Biernat:2017lfv}
K.~Golec-Biernat and S.~Sapeta,
JHEP \textbf{03}, 102 (2018)
doi:10.1007/JHEP03(2018)102
[arXiv:1711.11360 [hep-ph]].

\bibitem{Bartels:2002cj}
J.~Bartels, K.~J.~Golec-Biernat and H.~Kowalski,
Phys. Rev. D \textbf{66}, 014001 (2002)
doi:10.1103/PhysRevD.66.014001
[arXiv:hep-ph/0203258 [hep-ph]].

\bibitem{Kowalski:2003hm}
H.~Kowalski and D.~Teaney,
Phys. Rev. D \textbf{68}, 114005 (2003)
doi:10.1103/PhysRevD.68.114005
[arXiv:hep-ph/0304189 [hep-ph]].

\bibitem{Chachamis:2015ona}
G.~Chachamis, M.~De{\'a}k, M.~Hentschinski, G.~Rodrigo and A.~Sabio Vera,
JHEP \textbf{09}, 123 (2015)
doi:10.1007/JHEP09(2015)123
[arXiv:1507.05778 [hep-ph]].

\bibitem{Celiberto:2018muu}
F.~G.~Celiberto, D.~Gordo G{\'o}mez and A.~Sabio Vera,
Phys. Lett. B \textbf{786}, 201-206 (2018)
doi:10.1016/j.physletb.2018.09.045
[arXiv:1808.09511 [hep-ph]].

\bibitem{Cepila:2019skb}
J.~Cepila, J.~Nemchik, M.~Krelina and R.~Pasechnik,
Eur. Phys. J. C \textbf{79}, no.6, 495 (2019)
doi:10.1140/epjc/s10052-019-7016-9
[arXiv:1901.02664 [hep-ph]].

\bibitem{Peredo:2023oym}
M.~A.~Peredo and M.~Hentschinski,
Phys. Rev. D \textbf{109}, no.1, 014032 (2024)
doi:10.1103/PhysRevD.109.014032
[arXiv:2308.15430 [hep-ph]].

\bibitem{ALICE:2021tyx}
S.~Acharya \textit{et al.} [ALICE],
Phys. Lett. B \textbf{817}, 136280 (2021)
doi:10.1016/j.physletb.2021.136280
[arXiv:2101.04623 [nucl-ex]].

\bibitem{Brodsky:1980vj}
S.~J.~Brodsky, T.~Huang and G.~P.~Lepage,
SLAC-PUB-2540.

\bibitem{Cox:2009ag}
B.~E.~Cox, J.~R.~Forshaw and R.~Sandapen,
JHEP \textbf{06}, 034 (2009)
doi:10.1088/1126-6708/2009/06/034
[arXiv:0905.0102 [hep-ph]].

\bibitem{Nemchik:1994fp}
J.~Nemchik, N.~N.~Nikolaev and B.~G.~Zakharov,
Phys. Lett. B \textbf{341}, 228-237 (1994)
doi:10.1016/0370-2693(94)90314-X
[arXiv:hep-ph/9405355 [hep-ph]].

\bibitem{Kowalski:2006hc}
H.~Kowalski, L.~Motyka and G.~Watt,
Phys. Rev. D \textbf{74}, 074016 (2006)
doi:10.1103/PhysRevD.74.074016
[arXiv:hep-ph/0606272 [hep-ph]].

\bibitem{Armesto:2014sma}
N.~Armesto and A.~H.~Rezaeian,
Phys. Rev. D \textbf{90}, no.5, 054003 (2014)
doi:10.1103/PhysRevD.90.054003
[arXiv:1402.4831 [hep-ph]].

\bibitem{Goncalves:2014swa}
V.~P.~Gon{\c{c}}alves, B.~D.~Moreira and F.~S.~Navarra,
Phys. Lett. B \textbf{742}, 172-177 (2015)
doi:10.1016/j.physletb.2015.01.035
[arXiv:1408.1344 [hep-ph]].

\bibitem{Frankfurt:1996ri}
L.~Frankfurt, A.~Radyushkin and M.~Strikman,
Phys. Rev. D \textbf{55}, 98-104 (1997)
doi:10.1103/PhysRevD.55.98
[arXiv:hep-ph/9610274 [hep-ph]].

\bibitem{Mantysaari:2018nng}
H.~M{\"a}ntysaari and P.~Zurita,
Phys. Rev. D \textbf{98}, 036002 (2018)
doi:10.1103/PhysRevD.98.036002
[arXiv:1804.05311 [hep-ph]].

\bibitem{DeVries:1987atn}
H.~De Vries, C.~W.~De Jager and C.~De Vries,
Atom. Data Nucl. Data Tabl. \textbf{36}, 495-536 (1987)
doi:10.1016/0092-640X(87)90013-1

\bibitem{Cepila:2020uxc}
J.~Cepila and M.~Matas,
Eur. Phys. J. A \textbf{56}, no.9, 232 (2020)
doi:10.1140/epja/s10050-020-00243-4
[arXiv:2006.16136 [hep-ph]].

\bibitem{Deganutti:2023qct}
F.~Deganutti, C.~Royon and S.~Schlichting,
JHEP \textbf{01}, 159 (2024)
doi:10.1007/JHEP01(2024)159
[arXiv:2311.01965 [hep-ph]].

\bibitem{ZEUS:2002wfj}
S.~Chekanov \textit{et al.} [ZEUS],
Eur. Phys. J. C \textbf{24}, 345-360 (2002)
doi:10.1007/s10052-002-0953-7
[arXiv:hep-ex/0201043 [hep-ex]].

\bibitem{Chekanov:2004mw}
S.~Chekanov \textit{et al.} [ZEUS],
Nucl. Phys. B \textbf{695}, 3-37 (2004)
doi:10.1016/j.nuclphysb.2004.06.034
[arXiv:hep-ex/0404008 [hep-ex]].

\bibitem{H1:2005dtp}
A.~Aktas \textit{et al.} [H1],
Eur. Phys. J. C \textbf{46}, 585-603 (2006)
doi:10.1140/epjc/s2006-02519-5
[arXiv:hep-ex/0510016 [hep-ex]].

\bibitem{H1:2013okq}
C.~Alexa \textit{et al.} [H1],
Eur. Phys. J. C \textbf{73}, no.6, 2466 (2013)
doi:10.1140/epjc/s10052-013-2466-y
[arXiv:1304.5162 [hep-ex]].

\bibitem{Alexa:2013xxa}
C.~Alexa \textit{et al.} [H1],
Eur. Phys. J. C \textbf{73}, no.6, 2466 (2013)
doi:10.1140/epjc/s10052-013-2466-y
[arXiv:1304.5162 [hep-ex]].

\bibitem{Aktas:2005xu}
A.~Aktas \textit{et al.} [H1],
Eur. Phys. J. C \textbf{46}, 585-603 (2006)
doi:10.1140/epjc/s2006-02519-5
[arXiv:hep-ex/0510016 [hep-ex]].

\bibitem{ALICE:2018oyo}
S.~Acharya \textit{et al.} [ALICE],
Eur. Phys. J. C \textbf{79}, no.5, 402 (2019)
doi:10.1140/epjc/s10052-019-6816-2
[arXiv:1809.03235 [nucl-ex]].

\bibitem{ALICE:2023mfc}
S.~Acharya \textit{et al.} [ALICE],
Phys. Rev. D \textbf{108}, no.11, 112004 (2023)
doi:10.1103/PhysRevD.108.112004
[arXiv:2304.12403 [nucl-ex]].

\bibitem{Aaij:2018arx}
R.~Aaij \textit{et al.} [LHCb],
JHEP \textbf{10}, 167 (2018)
doi:10.1007/JHEP10(2018)167
[arXiv:1806.04079 [hep-ex]].

\bibitem{LHCb:2024pcz}
R.~Aaij \textit{et al.} [LHCb],
SciPost Phys. \textbf{18}, no.2, 071 (2025)
doi:10.21468/SciPostPhys.18.2.071
[arXiv:2409.03496 [hep-ex]].

\bibitem{H1:2000kis}
C.~Adloff \textit{et al.} [H1],
Phys. Lett. B \textbf{483}, 23-35 (2000)
doi:10.1016/S0370-2693(00)00530-X
[arXiv:hep-ex/0003020 [hep-ex]].

\bibitem{ZEUS:2009asc}
S.~Chekanov \textit{et al.} [ZEUS],
Phys. Lett. B \textbf{680}, 4-12 (2009)
doi:10.1016/j.physletb.2009.07.066
[arXiv:0903.4205 [hep-ex]].

\bibitem{LHCb:2015wlx}
R.~Aaij \textit{et al.} [LHCb],
JHEP \textbf{09}, 084 (2015)
doi:10.1007/JHEP09(2015)084
[arXiv:1505.08139 [hep-ex]].

\bibitem{CMS:2018bbk}
A.~M.~Sirunyan \textit{et al.} [CMS],
Eur. Phys. J. C \textbf{79}, no.3, 277 (2019)
[erratum: Eur. Phys. J. C \textbf{82}, no.4, 343 (2022)]
doi:10.1140/epjc/s10052-019-6774-8
[arXiv:1809.11080 [hep-ex]].

\bibitem{Guzey:2013xba}
V.~Guzey, E.~Kryshen, M.~Strikman and M.~Zhalov,
Phys. Lett. B \textbf{726}, 290-295 (2013)
doi:10.1016/j.physletb.2013.08.043
[arXiv:1305.1724 [hep-ph]].

\bibitem{Luszczak:2022fkf}
A.~{\L}uszczak, M.~{\L}uszczak and W.~Sch{\"a}fer,
Phys. Lett. B \textbf{835}, 137582 (2022)
doi:10.1016/j.physletb.2022.137582
[arXiv:2210.02877 [hep-ph]].

\bibitem{SabioVera:2006cza}
A.~Sabio Vera,
Nucl. Phys. B \textbf{746}, 1-14 (2006)
doi:10.1016/j.nuclphysb.2006.04.004
[arXiv:hep-ph/0602250 [hep-ph]].

\bibitem{SabioVera:2007ndx}
A.~Sabio Vera and F.~Schwennsen,
Nucl. Phys. B \textbf{776}, 170-186 (2007)
doi:10.1016/j.nuclphysb.2007.03.050
[arXiv:hep-ph/0702158 [hep-ph]].

\bibitem{Brodsky:1982gc}
S.~J.~Brodsky, G.~P.~Lepage and P.~B.~Mackenzie,
Phys. Rev. D \textbf{28}, 228 (1983)
doi:10.1103/PhysRevD.28.228

\bibitem{Brodsky:2002ka}
S.~J.~Brodsky, V.~S.~Fadin, V.~T.~Kim, L.~N.~Lipatov and G.~B.~Pivovarov,
JETP Lett. \textbf{76}, 249-252 (2002)
doi:10.1134/1.1520615
[arXiv:hep-ph/0207297 [hep-ph]].

\bibitem{Celiberto:2022dyf}
F.~G.~Celiberto and M.~Fucilla,
Eur. Phys. J. C \textbf{82}, no.10, 929 (2022)
doi:10.1140/epjc/s10052-022-10818-8
[arXiv:2202.12227 [hep-ph]].

\bibitem{Chachamis:2022jis}
G.~Chachamis and A.~Sabio Vera,
JHEP \textbf{07}, 109 (2022)
doi:10.1007/JHEP07(2022)109
[arXiv:2203.12418 [hep-th]].
\end{thebibliography}

\end{document}